\documentclass[aip, jcp, preprint,showpacs,superscriptaddress,groupedaddress]{revtex4-1}

\usepackage{amsmath}
\usepackage{amssymb}
\usepackage{braket}
\usepackage{bm}

\usepackage{graphicx}
\usepackage{subcaption}

\begin{document}

\title{Density Matrices of Seniority-Zero Geminal Wavefunctions}
\author{Jean-David Moisset}
\author{Charles-\'{E}mile Fecteau}
\author{Paul A. Johnson}
 \email{paul.johnson@chm.ulaval.ca}
 \affiliation{D\'{e}partement de chimie, Universit\'{e} Laval, Qu\'{e}bec, Qu\'{e}bec, G1V 0A6, Canada}

\date{\today}

\begin{abstract}
Scalar products and density matrix elements of closed-shell pair geminal wavefunctions are evaluated directly in terms of the pair amplitudes, resulting in an analogue of Wick's theorem for fermions or bosons. This expression is in general intractable, but it is shown how it becomes feasible in three distinct ways for Richardson-Gaudin (RG) states, the antisymmetrized geminal power, and the antisymmetrized product of strongly-orthogonal geminals. Dissociation curves for hydrogen chains are computed with off-shell RG states and the antisymmetrized product of interacting geminals. Both are near exact suggesting that the incorrect results observed with ground state RG states are fixable using a different RG state.
\end{abstract}

\maketitle

\section{Introduction}
Weakly-correlated systems are well-described in the orbital picture. The physical wavefunction is dominated by the Hartree-Fock (HF) Slater determinant of the occupied orbitals, with small contributions from Slater determinants corresponding to single and double excitations. The qualitative physical behaviour is weakly-interacting electrons. Kohn-Sham Density Functional Theory (DFT) and Coupled-Cluster (CC) with singles and doubles treat weakly-correlated systems quite well.\cite{helgaker_book}

Strongly-correlated systems are much more difficult to treat in the orbital picture. There is not one, but many important Slater determinants along with their corresponding single- and double-excitations which must be included for a correct treatment. If only a few important Slater determinants are required, approaches such as the complete active space self-consistent field (CASSCF) or complete active space configuration interaction (CASCI) are good treatments, but these become intractable as the number of important Slater determinants grows. It is not always easy to identify the important Slater determinants by hand. Sophisticated algorithms such as the Density Matrix Renormalization Group (DMRG),\cite{white:1992,white:1993,chan:2002,chan:2004,chan:2011} Slater Determinant Monte-Carlo (SDMC),\cite{thom:2005,booth:2010,booth:2013} Configuration Interaction using a Perturbative Selection made Iteratively (CIPSI)\cite{huron:1973} and Semistochastic Heat-Bath Configuration Interaction (SHCI)\cite{sharma:2017,holmes:2017,li:2018,yao:2021} are able to treat strongly-correlated systems well by efficiently evaluating high-rank expressions or efficiently picking the important Slater determinants.

It has long been understood that two-electron building blocks, geminals, are a better starting point for strongly-correlated systems.\cite{fock:1950,mcweeny:1959,mcweeny:1960,mcweeny:1963} Indeed, as the Coulomb interaction is a two-electron interaction, a picture of weakly-interacting two-electron objects should be a good starting point. However, unless structure is imposed, geminal product wavefunctions are computationally intractable. The most general geminal wavefunction possible, the antisymmetrized product of geminals (APG)\cite{nicely:1971,siems:1976} is known to give good results, but is algebraically very difficult to work with. Restriction to closed-shell pairs gives the antisymmetrized product of interacting geminals (APIG),\cite{silver:1969,silver:1970a,silver:1970b,silver:1970c} which is still not feasible. Further simplifications are required. The present contribution will focus on the scalar products and density matrix elements of APIG and its descendants with a view to more general cases in the future. This treatment of APIG is a stepping-stone to a similar treatment of APG. 

APIG is a seniority-zero wavefunction, which is to say there are zero unpaired electrons. Seniority-zero wavefunctions are built from a pair representation of the Lie algebra su(2). APIG has been observed to be quite a good approximation to doubly-occupied configuration interaction (DOCI),\cite{weinhold:1967a,weinhold:1967b,cook:1975} the most general seniority-zero wavefunction. DOCI describes single-bond dissociation processes near exactly, and has been shown to be a reasonable starting point for more general dissociations.\cite{bytautas:2011} 

Many degenerate cases of APIG are known by different names in different fields.\cite{johnson:2013} The algebraic Bethe Ansatz (ABA)\cite{bethe:1931,faddeev:1980,korepin_book} solutions to the reduced Bardeen-Cooper-Schrieffer (BCS)\cite{bardeen:1957a,bardeen:1957b} Hamiltonian, which we call Richardson-Gaudin (RG)\cite{richardson:1963,richardson:1964,richardson:1965,gaudin:1976} states, are a case of APIG for which the geminal coefficients are parametrized by a set of complex numbers that satisfy a set of non-linear equations. The antisymmetrized geminal power (AGP)\cite{coleman:1965} is a case of APIG for which the geminals are the same. It is possible to define a more general AGPs, but all can be reduced to closed-shell pairs in a particular basis. AGP is also known in the condensed matter literature as a number-projected BCS wavefunction, and in the nuclear structure literature as a number-projected coherent state. The antisymmetrized product of strongly-orthogonal geminals (APSG)\cite{hurley:1953,kutzelnigg:1964} is an APIG for which the geminals act on distinct sets of spatial orbitals. In the  antisymmetrized product of 1-reference orbital geminals (AP1roG),\cite{limacher:2013} the spatial orbitals are separated into occupied and virtuals with each geminal receiving contributions from a single occupied spatial orbital but each virtual spatial orbital. AP1roG is equivalent to pair coupled-cluster doubles (pCCD)\cite{stein:2014} and is solved by projection.

Recently, we have presented variational results using the RG ground state as a mean-field wavefunction. For symmetric dissociations of hydrogen chains, a prototypical strongly-correlated system well described by DOCI, we observed qualitatively incorrect results. Frankly, this was rather disappointing. The motivation for the present work was to see whether this qualitatively incorrect behaviour was the fault of the RG geminal form, or of APIG in general. Calculations performed with AP1roG/pCCD show near perfect agreement with DOCI for these systems,\cite{limacher:2013} so APIG is certainly correct. We therefore set out to perform variational calculations with RG states with arbitrary geminal coefficients, which are known as off-shell RG states. This led us to calculate the scalar products and density matrix elements for off-shell RG states and for APIG.

Density matrix elements for APIG are known, and have been calculated by expanding APIG in a basis of Slater determinants. This approach clashes with the known results for the density matrix elements of RG states which are calculated by using the structure of the pairs themselves. The single-electron basis information appears only in complete summations. We follow a similar approach for APIG which, as far as we know, has not been done. The scalar product between two APIG states becomes a sum over possible ways of contracting the geminal coefficients, which can be seen as a higher-rank analogue of Wick's theorem for fermions or bosons. The shape of the result is identical to one obtained by Sklyanin\cite{sklyanin:1999} for off-shell RG states, so we refer to it as a Sklyanin sum. For APIG the Sklyanin sum is intractable to evaluate, but it becomes numerically feasible in three different ways. For off-shell RG the rank of the geminal coefficient contractions is reducible, for AGP the contractions depend only on their rank leading to recursion, and for APSG the non-zero contractions are sparse. Going forward, we will use a similar approach for APG as one or more of the present reductions in complexity will simplify the analogue of the corresponding Sklyanin sum.

In section \ref{sec:senzero} the algebraic structure of seniority-zero wavefunctions is summarized, focusing on RG states. Scalar products and density matrix elements for APIG are calculated directly in terms of the geminal coefficients in section \ref{sec:apig}, with its degenerate cases following in section \ref{sec:offshell} for off-shell RG states, section \ref{sec:agp} for AGP and section \ref{sec:apsg} for APSG. Symmetric hydrogen chain dissociation curves are calculated numerically, for off-shell RG and APIG, in section \ref{sec:numbers}. As APIG scalar products and density matrix elements are known in a basis of Slater determinants, each expression has been verified numerically. As always, while the intermediate stages of the development may be tricky and tedious, we aim to make our final expressions as simple as possible. The algebraic message of the Sklyanin sum is clear, and the final results are not complicated. This contribution is not a review, but is meant to tie together many similar results from different fields. 

\section{Closed-Shell pairs: su(2)}\label{sec:senzero}
Pairs are built with the objects
\begin{align} \label{eq:su2_objects}
S^+_i = a^{\dagger}_{i\uparrow} a^{\dagger}_{i\downarrow}, \quad S^-_i = a_{i\downarrow} a_{i\uparrow}, \quad
S^z_i = \frac{1}{2} \left( a^{\dagger}_{i\uparrow}a_{i\uparrow} + a^{\dagger}_{i\downarrow}a_{i\downarrow} -1 \right)
\end{align}
where $a^{\dagger}_{i\uparrow}$ creates an up-spin electron in spatial orbital $i$ etc. $S^+_i$ creates a pair of electrons in the spatial orbital $i$ while $S^-_i$ removes a pair from spatial orbital $i$. It is useful to use $\hat{n}_i = 2S^z_i + 1$ which counts the number of pairs in spatial orbital $i$. The objects \eqref{eq:su2_objects} have su(2) structure constants
\begin{subequations} \label{eq:su2_structure}
\begin{align} 
[S^+_i , S^-_j] &= 2 \delta_{ij} S^z_i \\
[S^z_i , S^{\pm}_j] &= \pm \delta_{ij} S^{\pm}_i.
\end{align}
\end{subequations}
All seniority-zero states are built with these objects. In this particular representation the pairs are all localised up-down spin partners. More general pairing schemes could be engineered,\cite{johnson:2017} though for clarity we will keep the representation \eqref{eq:su2_objects}. The vacuum $\ket{\theta}$ is destroyed by each $S^-_i$ and is an eigenvector of each $S^z_i$
\begin{subequations} \label{eq:su2vac}
\begin{align}
S^-_i \ket{\theta} &= 0 \\
S^z_i \ket{\theta} &= -\frac{1}{2} \ket{\theta}.
\end{align}
\end{subequations}
Typically this is the physical vacuum, though any Slater determinant of unpaired electrons is allowable. The only requirement is that they do not participate in the pairing scheme.\cite{coleman_book} 

There is also the $\frac{1}{2}-$spin representation of su(2)
\begin{align} \label{eq:su2_objects_spin}
\tilde{S}^+_i = a^{\dagger}_{i\uparrow}a_{i\downarrow}, \quad \tilde{S}^-_i = a^{\dagger}_{i\downarrow}a_{i\uparrow}, \quad
\tilde{S}^z_i = \frac{1}{2} \left( a^{\dagger}_{i\uparrow}a_{i\uparrow} - a^{\dagger}_{i\downarrow}a_{i\downarrow} \right)
\end{align}
which has the same structure constants \eqref{eq:su2_structure}. In this representation, the vacuum is the ferromagnetic state with all spins aligned downward. All the results in the pair representation are immediately transferable to the $\frac{1}{2}-$spin representation though we will not consider it any further.

Slater determinants are labelled with sets of indices $\{i\}$ based on which spatial orbitals are occupied. All the Slater determinants in this paper are doubly-occupied, and hence are written
\begin{align} \label{eq:slater_notation}
\ket{\{i\}} = S^+_{i_1} S^+_{i_2} \dots S^+_{i_M}\ket{\theta}.
\end{align}
While this notation may seem cumbersome, it serves to emphasize that \emph{Slater determinants are not the optimal basis for geminal wavefunctions}. Without exception, all states represent $M$ pairs among $N$ spatial orbitals. The basis of Slater determinants therefore contains $\binom{N}{M}$ elements. Generally, $i,j,k,l$ will label spatial orbitals while $a,b,c,d$ will label pairs.

The physical Hamiltonian we wish to solve is the molecular Coulomb Hamiltonian
\begin{align} \label{eq:C_ham}
\hat{H}_C = \sum_{ij} h_{ij} \sum_{\sigma} a^{\dagger}_{i \sigma} a_{j \sigma} + \frac{1}{2} \sum_{ijkl} V_{ijkl} \sum_{\sigma \tau} a^{\dagger}_{i \sigma} a^{\dagger}_{j \tau} a_{l \tau} a_{k \sigma}
\end{align}
for which the summations over $\sigma$ and $\tau$ represent the two components of the spin, and the 1- and 2-electron integrals
\begin{align}
h_{ij} &= \int d\mathbf{r} \phi^*_i (\mathbf{r}) \left( - \frac{1}{2} \nabla^2 - \sum_I \frac{Z_I}{| \mathbf{r} - \mathbf{R}_I |} \right) \phi_j (\mathbf{r}) \\
V_{ijkl} &= \int d\mathbf{r}_1 d\mathbf{r}_2 \frac{\phi^*_i(\mathbf{r}_1)  \phi^*_j(\mathbf{r}_2)  \phi_k(\mathbf{r}_1)  \phi_l(\mathbf{r}_2)  }{| \mathbf{r}_1 - \mathbf{r}_2|}
\end{align}
are pre-computed in a basis of single-particle functions $\{ \phi \}$. We will henceforth assume that they are real, though no substantial complication arises if they are not. 

States built from the pair creators $\eqref{eq:su2_objects}$ will have zero seniority. Generally, the physical wavefunction has contributions with any number of unpaired electrons, so seniority-zero wavefunctions are only an approximation. However, solving the seniority-zero problem is already a difficult task, and itself represents a stepping-stone to more general structures. It has been demonstrated that even for the dissociation for molecular nitrogen, a problem with several important seniority sectors, solving the seniority-zero problem exactly is already a reasonable starting point.\cite{bytautas:2011}  

With a seniority-zero state $\ket{\Psi}$, the only terms in the Coulomb Hamiltonian \eqref{eq:C_ham} that contribute are
\begin{align} \label{eq:sen0_ham}
\hat{H}_{SZ} = \sum_i h_{ii} \hat{n}_i + \frac{1}{4} \sum_{i \neq j} \left(2V_{ijij} - V_{ijji} \right) \hat{n}_i \hat{n}_j + \sum_{ij} V_{iijj} S^+_i S^-_j.
\end{align}
The second summation is a double summation over $i$ and $j$ such that the diagonal elements are left out. The Hamiltonian \eqref{eq:sen0_ham} leaves out the terms of \eqref{eq:C_ham} with non-zero seniorities, and is hence \emph{not} invariant to orbital transformations. Optimal orbitals for seniority-zero wavefunctions are known to be localized.\cite{limacher:2014a} To evaluate the expected value $\braket{\Psi | \hat{H}_{SZ} | \Psi}$ the only matrix elements required are 
\begin{subequations} \label{eq:sz_dm}
\begin{align}
\gamma_k &= \frac{1}{2} \frac{\braket{\Psi | \hat{n}_k | \Psi}}{\braket{\Psi | \Psi}} \\
D_{kl} &= \frac{1}{4}   \frac{\braket{\Psi | \hat{n}_k \hat{n}_l | \Psi}} {\braket{\Psi | \Psi}} \\
P_{kl} &= \frac{\braket{\Psi | S^+_k S^-_l | \Psi}} {\braket{\Psi | \Psi}}.
\end{align}
\end{subequations}
The 1-body reduced density matrix $\gamma_k$ is diagonal, while the 2-body reduced density matrix has two non-zero pieces: the \emph{diagonal-correlation function} $D_{kl}$ and the \emph{pair-correlation function} $P_{kl}$. Note that the diagonal term of each refers to the same element $D_{kk} = P_{kk}$ and further $P_{kk} = \gamma_k$. The Hamiltonian \eqref{eq:sen0_ham} avoids double-counting as $D_{kk}$ is not included, so as a convention we set $D_{kk}=0$. We will refer to the objects \eqref{eq:sz_dm} as 1- and 2-body density matrix elements (DM) as most of the results apply equally well to reduced density matrix elements (for one seniority-zero state) and to transition density matrix elements (between distinct seniority-zero states). There are sum rules for the diagonal elements:
\begin{align}
\sum_k \gamma_k &= M \\
\sum_{kl} D_{kl} &= M(M-1).
\end{align}

The best possible seniority-zero wavefunction is doubly-occupied configuration interaction (DOCI).\cite{weinhold:1967a,weinhold:1967b,cook:1975} The variables are the coefficients $C^{\{i\}}$ of a CI expansion in doubly-occupied Slater determinants:
\begin{align} \label{eq:DOCI}
\ket{\text{DOCI}} = \sum_{\{i\}} C^{\{i\}} \ket{\{i\}}.
\end{align}
There are $\binom{N}{M}$ Slater determinants and thus $\binom{N}{M}$ coefficients. The summation in \eqref{eq:DOCI} should be understood as over all Slater determinants \eqref{eq:slater_notation}. Written explicitly, the summation would be over all indices $i_1,\dots, i_M$ such that $i_1 < i_2 < \dots < i_M$. DOCI is not invariant to orbital transformation, and thus numerical comparisons must be made to orbital-optimized (OO)-DOCI. At present our goal is to represent DOCI effectively with geminal products so the Hamiltonian \eqref{eq:sen0_ham} will always be assumed to be written in the OO-DOCI orbitals.

Recently\cite{johnson:2020,fecteau:2020,fecteau:2021,johnson:2021} we have employed the eigenvectors of the reduced Bardeen-Cooper-Schrieffer (BCS) Hamiltonian\cite{bardeen:1957a,bardeen:1957b}
\begin{align} \label{eq:hbcs}
\hat{H}_{BCS} = \frac{1}{2} \sum_i \varepsilon_i \hat{n}_i - \frac{g}{2} \sum_{ij} S^+_i S^-_j,
\end{align}
the so-called isotropic Richardson-Gaudin (RG)\cite{richardson:1963,richardson:1964,richardson:1965,gaudin:1976} states, as a mean-field wavefunction. With the Lie algebra of pair objects
\begin{align}
S^{\pm}(u) = \sum_i \frac{S^{\pm}_i}{u - \varepsilon_i}, \quad S^z(u) = \frac{1}{g} - \sum_i \frac{S^z_i}{u- \varepsilon_i},
\end{align}
which satisfy the structure
\begin{align}
[S^+(u),S^-(v)] &= 2 \frac{S^z(u) - S^z(v)}{u-v} \\
[S^z(u),S^{\pm}(v)] &= \pm \frac{S^{\pm}(u) - S^{\pm} (v)}{u-v}
\end{align}
the RG states are 
\begin{align} \label{eq:rg_off}
\ket{\{u\}} = S^+(u_1) S^+(u_2) \dots S^+(u_M) \ket{\theta}.
\end{align}
The states \eqref{eq:rg_off} are eigenvectors of \eqref{eq:hbcs} provided that the complex numbers $\{u\}$, the \emph{rapidities}, are solutions of the set of coupled non-linear equations
\begin{align} \label{eq:rich_eq}
\frac{2}{g} + \sum_i \frac{1}{u_a - \varepsilon_i} + \sum_{b(\neq a)} \frac{2}{u_b - u_a} = 0, \qquad \forall a = 1,\dots M.
\end{align}
The second summation is a single summation over $b$ with the $a$th element left out. This notation is to be contrasted with \eqref{eq:sen0_ham} where it is a double sum. Richardson's equations \eqref{eq:rich_eq} ensure that the action of \eqref{eq:hbcs} on \eqref{eq:rg_off} yields only a single term proportional to \eqref{eq:rg_off}. This is a particular case of the algebraic Bethe ansatz (ABA)\cite{bethe:1931,faddeev:1980,korepin_book} construction (see refs.\cite{carrier:2020,moisset:2021} for an introduction in terms of individual electrons).  The eigenvalue problem for \eqref{eq:hbcs} has been reduced to a set of coupled non-linear equations to solve for the rapidities $\{u\}$. This yields not one, but \emph{all} the eigenvectors of \eqref{eq:hbcs}, though as Richardson's equations are coupled, each eigenvector is parametrized by a distinct set of rapidities. Simple particle-hole type excitations with second-quantized operators from one state to another are not possible for RG states. Richardson's equations must be solved numerically for which many algorithms exist,\cite{rombouts:2004,guan:2012,pogosov:2012,debaerdemacker:2012,claeys:2015} with the most efficient being that of refs.\cite{faribault:2011,elaraby:2012}

Without exception, rapidities emphasized with a tilde $\{ \tilde{v}\}$ will always represent solutions of Richardson's equations while $\{u\}$ are arbitrary complex numbers. Solutions of Richardson's equations $\{ \tilde{v}\}$ are \emph{on-shell} while arbitrary $\{u\}$ are \emph{off-shell}. Ordinarily we would emphasize which set of rapidities were arbitrary, but as the main focus of this contribution is scalar products for more general su(2) geminal wavefunctions, we will emphasize the sets that are on-shell with tildes.

The machinery behind the DM elements for on-shell RG states is known, and has been presented many times,\cite{amico:2002,faribault:2008,faribault:2010,fecteau:2020} so we will very quickly outline the approach. The simplest final expressions are results of Gorohovsky and Bettelheim.\cite{gorohovsky:2011} The starting point is Slavnov's theorem\cite{slavnov:1989,zhou:2002} for the scalar product of an on-shell RG state with rapidities $\{\tilde{v}\}$ and an off-shell RG state with rapidities $\{u\}$:
\begin{align} \label{eq:slavnov}
\braket{ \{\tilde{v} \} | \{ u \}} = K (\{\tilde{v}\},\{u\} ) \det J( \{\tilde{v}\}, \{u\} )
\end{align}
with
\begin{align}
K (\{\tilde{v}\},\{u\} )  = \frac{\prod_{ab} (\tilde{v}_a-u_b)}{\prod_{a<b} (u_a-u_b)(\tilde{v}_b - \tilde{v}_a)}
\end{align}
and
\begin{align}
J_{ab} = \frac{1}{(\tilde{v}_a-u_b)^2} \left( 2 \alpha(u_b) - \sum_{c (\neq a)} \frac{2}{(u_b - u_c)} \right).
\end{align}
Notice that in the matrix $J$ the only information required are the rapidities $\{u\}$ and the vacuum eigenvalues $\alpha(u)$,
\begin{align}
S^z(u) \ket{\theta} &= \alpha(u) \ket{\theta} \\
\alpha(u) &= \frac{1}{g} + \frac{1}{2} \sum_i \frac{1}{u - \varepsilon_i}
\end{align}
which are both properties of the pairs. The single particle energies of the reduced BCS Hamiltonian $\{\varepsilon\}$ appear only in a complete summation. 

Taking the limit $\{u\} \rightarrow \{ \tilde{v} \}$ gives the square of the norm as the determinant of the \emph{Gaudin} matrix
\begin{align}
\braket{ \{\tilde{v} \} | \{ \tilde{v} \}} = \det G
\end{align}
which is understood as the Jacobian of Richardson's equations:
\begin{align} \label{eq:gmat}
G_{ab} = 
\begin{cases}
\sum_i \frac{1}{(\tilde{v}_a - \varepsilon_i)^2} -\sum_{c \neq a} \frac{2}{( \tilde{v}_a - \tilde{v}_c )^2} , &\quad a = b \\
\frac{2}{(\tilde{v}_a - \tilde{v}_b )^2}, &\quad a \neq b.
\end{cases}
\end{align}

The 1-DM elements are obtained by moving $\hat{n}_k$ past each pair creator $S^+(u)$, using
\begin{align}
[\hat{n}_k , S^+(u)] = \frac{S^+_k}{u - \varepsilon_k}
\end{align}
until $\hat{n}_k$ destroys the vacuum. The result is a sum of scalar products with one $u$ missing, called \emph{form factors}, which are easily evaluated as limiting cases of Slavnov's theorem. Specifically,
\begin{align} \label{eq:di_ff}
\frac{1}{2} \braket{ \{ \tilde{v} \} | \hat{n}_k | \{ u \}} = \sum_a \frac{\braket{ \{ \tilde{v} \} | S^+_k | \{u\}_a }}{(u_a - \varepsilon_k)}
\end{align}
where $\braket{ \{ \tilde{v} \} | S^+_k | \{u\}_a }$ is a form factor and $\{u\}_a$ is the set $\{u\}$ without the element $u_a$. Given \eqref{eq:slavnov}, we can evaluate these scalar products provided that we know how to write the local pair creator $S^+_k$ in terms of $S^+(u)$. In the ABA this is known as the \emph{inverse problem}. For RG, the local objects are residues of the RG pairs at their simple poles:
\begin{align}
S^+_k = \lim_{u \rightarrow \varepsilon_k} (u - \varepsilon_k) S^+(u),
\end{align}
and therefore the form factor is the residue of the scalar product
\begin{align}
\braket{ \{ \tilde{v} \} | S^+_k | \{u\}_a } = \lim_{u_a \rightarrow \varepsilon_k} (u_a - \varepsilon_k)  \braket{ \{\tilde{v} \} | \{u \} }.
\end{align}
Evaluating the residue, then taking the limit $\{u\} \rightarrow \{ \tilde{v} \}$ gives
\begin{align}
\braket{ \{ \tilde{v} \} | S^+_k | \{ \tilde{v} \}_a } = (\tilde{v}_a - \varepsilon_k) \det G^k_a.
\end{align}
The matrix $G^k_a$ is the Gaudin matrix \eqref{eq:gmat} with the $k$th column replaced with the vector
\begin{align}
\textbf{b}_k =
\begin{pmatrix}
\frac{1}{( \tilde{v}_1 - \varepsilon_k ^2)} \\
\vdots \\
\frac{1}{( \tilde{v}_M - \varepsilon_k ^2)}
\end{pmatrix}.
\end{align}
The \emph{normalized} 1-RDM elements are then
\begin{align}
\gamma_k = \sum_a \frac{\det G^k_a}{\det G}
\end{align}
in which the primitive summands are easily obtained, from Cramer's rule, as the solutions of a system of linear equations. A brief physical argument shows that this is the same set of linear equations as for the partial derivatives of the rapidities with respect to the single particle energies.\cite{gorohovsky:2011} So, from the solutions of the linear equations
\begin{align} \label{eq:on_shell_primitives}
G \frac{\partial \tilde{\textbf{v}}}{\partial \varepsilon_k} = \textbf{b}_k
\end{align}
the 1-RDM elements are:
\begin{align}
\gamma_k = \sum_a \frac{\partial \tilde{v}_a}{\partial \varepsilon_k}.
\end{align}
The same procedure leads to clean expressions for \emph{normalized} $D_{kl}$ and $P_{kl}$. As they will be pertinent in the next section, we emphasize the un-normalized DM expressions in terms of form factors
\begin{align}
\frac{1}{4} \braket{ \{\tilde{v}\} | \hat{n}_k \hat{n}_l | \{u\}} &= \sum_{a\neq b} \frac{\braket{ \{\tilde{v}\} | S^+_k S^+_l | \{u\}_{a,b}  }}{(u_a - \varepsilon_k)(u_b - \varepsilon_l)}  \label{eq:Dij_ff} \\
\braket{ \{ \tilde{v} \} | S^+_k S^-_l | \{u\}} &= \sum_a \frac{\braket{ \{\tilde{v}\} | S^+_k | \{u\}_a }}{(u_a - \varepsilon_l)} 
- \sum_{a\neq b} \frac{\braket{ \{\tilde{v} \} | S^+_k S^+_l | \{u\}_{a,b}  }}{(u_a-\varepsilon_l)(u_b - \varepsilon_l)} \label{eq:Pij_ff}
\end{align}
where $\{u\}_{a,b}$ is the set $\{u\}$ without $u_a$ and $u_b$. The only additional result required is a relationship between scaled determinants:
\begin{align} \label{eq:jacobi}
\frac{\det G^{kl}_{ab}}{\det G} = \frac{\det G^k_a}{\det G} \frac{\det G^l_b}{\det G} - \frac{\det G^l_a}{\det G} \frac{\det G^k_b}{\det G}.
\end{align}
In \eqref{eq:jacobi}, the determinant of the matrix $G$ with two columns replaced, scaled by the original determinant $\det G$, is the same as a $2\times2$ determinant of scaled single-column replacements. This result is quite similar to a result of Jacobi,\cite{vein_book} and extends to any order: the scaled determinant of a matrix with $p$ replaced columns is a $p\times p$ determinant of scaled single-column updates. We will not prove this, but inform the interested reader that it falls out naturally from the rank-$p$ version of the matrix determinant lemma for the $N \times N$ matrix $A$
\begin{align}
\det ( A + UV^T) = \det (I + V^T A^{-1} U) \det(A)
\end{align}
using the explicit inverse formula in terms of the adjugate
\begin{align}
A^{-1} = \frac{\text{adj}(A)}{\det A}.
\end{align}
$U$ is the $N\times p$ matrix of 0s and 1s labelling which elements are updated, and $V^T$ a $p \times N$ matrix with the explicit updates.

For $k \neq l$, the 2-RDM elements are
\begin{align}
D_{kl} &= \sum_{a<b} \frac{(\tilde{v}_a - \varepsilon_k)(\tilde{v}_b-\varepsilon_l) + (\tilde{v}_a - \varepsilon_l)(\tilde{v}_b-\varepsilon_k)}{(\varepsilon_k - \varepsilon_l) (\tilde{v}_b-\tilde{v}_a)} \left( \frac{\partial \tilde{v}_a}{\partial \varepsilon_k}  \frac{\partial \tilde{v}_b}{\partial \varepsilon_l} -  \frac{\partial \tilde{v}_a}{\partial \varepsilon_l} \frac{\partial \tilde{v}_b}{\partial \varepsilon_k} \right) \\
P_{kl} &= \sum_a \frac{\tilde{v}_a - \varepsilon_k}{\tilde{v}_a - \varepsilon_l} \frac{\partial \tilde{v}_a}{\partial \varepsilon_k}
-2 \sum_{a<b} \frac{(\tilde{v}_b - \varepsilon_k)(\tilde{v}_a-\varepsilon_k) }{(\varepsilon_k - \varepsilon_l) (\tilde{v}_b-\tilde{v}_a)}
\left( \frac{\partial \tilde{v}_a}{\partial \varepsilon_k}  \frac{\partial \tilde{v}_b}{\partial \varepsilon_l} -  \frac{\partial \tilde{v}_a}{\partial \varepsilon_l} \frac{\partial \tilde{v}_b}{\partial \varepsilon_k} \right),
\end{align}
which were first obtained by Gorohovsky and Bettelheim\cite{gorohovsky:2011}. To evaluate the 2-RDM elements, all that is required is the primitives $\frac{\partial \tilde{v}_a}{\partial \varepsilon_k}$ which are obtained from solutions of the linear equations \eqref{eq:on_shell_primitives}.

In the next section we will look at APIG. Expressions for the scalar product and DM elements are known as complete summations in a basis of Slater determinants. We will evaluate them in a manner such that the single particle information, the individual geminal coefficients, appear only in complete summations as in Slavnov's theorem. Scalar products will be evaluated with the form factor approach. First, the scalar product between two arbitrary states will be computed. Next, the DM elements will be written as sums of form factors, which are each evaluated as limits of the scalar product with the solution of the corresponding inverse problem.

\section{APIG} \label{sec:apig}
APIG is a wavefunction first written by Silver\cite{silver:1969} as a generalization of AGP and APSG. While Silver was able to write first-quantized expressions for APIG's reduced density matrices (for two pairs), the expressions quickly become unmanageable. We will label APIG states with sets of geminal coefficients $\{g\},\{h\}$. They are the action of arbitrary closed-shell pair creators:
\begin{align} \label{eq:APIG_gem}
G^+(g_a) = \sum_i g^i_a S^+_i 
\end{align}
on the vacuum
\begin{align}
\ket{ \{g\}} = G^+(g_1)G^+(g_2) \dots G^+(g_M) \ket{\theta}.
\end{align} 
We choose to write the orbital index as a superscript (it has removed much confusion). We will emphasize that there is not a single coefficient $g_a$, but $N$ as there is a coefficient $g^i_a$ for each spatial orbital. It is useful to refer to the collection of geminal coefficients for the $a$th APIG geminal as $g_a$.

\subsection{Slater determinant expansion}
Expansion in Slater determinants is not difficult, and leads to large sums over permanents. Expressions for the DM elements computed in a Slater determinant basis are very easily obtained from the expressions of Weinhold and Wilson\cite{weinhold:1967a,weinhold:1967b} for DOCI. We will derive the same results with the form factor approach for two reasons: first to show how it works, and second to provide intermediate consistency checks for the results of the next section.

Resolving the identity in a basis of Slater determinants, we obtain
\begin{align} \label{eq:apig_ket_sd}
\ket{\{g\}} = \left( \sum_{ \{i\} } \ket{\{i\}} \bra{ \{i\}} \right) \ket{ \{g\}} =  \sum_{ \{i\} } C^{\{i\}}_{\{g\}} \ket{\{i\} }.
\end{align}
The expansion coefficients $C^{\{i\}}_{\{g\}} \equiv \braket{\{i\} | \{g\}}$ are permanents, or symmetric sums over the geminal coefficients:
\begin{align}
C^{\{i\}}_{\{g\}} = \underset{a,b} {\text{per}} \; \left(g^{i_b}_a \right) = 
\sum_{\sigma \in \mathfrak{S}_M} g^{i_{\sigma(1)}}_1 g^{i_{\sigma(2)}}_2 \dots g^{i_{\sigma(M)}}_M =
\begin{vmatrix}
g^{i_1}_1 & g^{i_2}_1 & \dots & g^{i_M}_1 \\
g^{i_1}_2 & g^{i_2}_2 & \dots & g^{i_M}_2 \\
\vdots & \vdots & \ddots & \vdots \\
g^{i_1}_M & g^{i_2}_M & \dots & g^{i_M}_M 
\end{vmatrix}^+.
\end{align}
The summation is over all permutations $\sigma$ in the symmetric group on $M$ elements $\mathfrak{S}_M$. Permanents are symmetric multi-linear functions: Laplace expansion may be performed along any row or column (or higher rank co-factors), the distinction being that all signs are positive.\cite{minc_book} Permanents are however \emph{not} invariant to row or column operations and are thus intractable to compute in general. The best known computational approach is due to Ryser,\cite{ryser_1963} which still does not scale polynomially. Like \eqref{eq:apig_ket_sd},
\begin{align}
\bra{\{h\}} = \sum_{\{j\}} C^{\{h\}}_{\{j\}} \bra{\{j\}}
\end{align}
so that the scalar product is
\begin{align} 
\braket{ \{h\} | \{g\} } &= \sum_{\{i\}} \sum_{\{j\}} C^{\{h\}}_{\{j\}} C^{\{i\}}_{\{g\}} \braket{ \{j\} | \{i\} } \\
&= \sum_{\{i\}} C^{\{h\}}_{\{i\}} C^{\{i\}}_{\{g\}} \label{eq:apig_slavnov_sd}
\end{align}
since the Slater determinants $\ket{\{i\}}$ form an orthonormal basis. The expression \eqref{eq:apig_slavnov_sd} is the master scalar product, in the basis of Slater determinants, that plays the role of Slavnov's theorem for on-shell RG. Expressions for RDMs are obtained as sums of form factors, which are obtained as specific cases. First notice that 
\begin{align}
[\hat{n}_k , G^+(g_a)] = 2 g^k_a
\end{align}
so that $\hat{n}_k$ can be moved to the right until it destroys the vacuum. Thus, in parallel to the form factor approach for RG we have
\begin{align}
\frac{1}{2}\braket{\{h\} | \hat{n}_k | \{g\}} &= \sum_a g^k_a \braket{ \{h\} | S^+_k | \{g\}_a} \\
\frac{1}{4}\braket{ \{h\} | \hat{n}_k \hat{n}_l | \{g\}} &= \sum_{a\neq b} g^k_a g^l_b \braket{ \{h\} | S^+_k S^+_l | \{g\}_{a,b}} \\
\braket{ \{h\} | S^+_k S^-_l | \{g\}} &= \sum_a g^l_a \braket{ \{h\} | S^+_k | \{g\}_a} - \sum_{a\neq b} g^l_a g^l_b \braket{ \{h\} | S^+_k S^+_l | \{g\}_{a,b}}
\end{align} 
where $\ket{\{g\}_a}$ is the $M-1$ pair APIG state for which the $a$th geminal coefficients have been removed, etc. These expressions are valid specifically when $k \neq l$. 

The form factors may be evaluated as the inverse problem for the APIG geminals \eqref{eq:APIG_gem} has solution
\begin{align}
S^+_k = \frac{\partial}{\partial g^k_a} G^+(g_a).
\end{align}

As permanents are multilinear functions, their derivatives are easy to compute. In general, derivatives of permanents would be sums over permanents with single rows (or columns) replaced with corresponding derivatives. Here, each permanent depends \emph{at most} linearly with respect to each parameter, so the derivative with respect to $g^k_a$ is the specific $(M-1)\times (M-1)$ minor of the permanent proportional to $g^k_a$:
\begin{align} \label{eq:1_der_apig_sd}
\frac{\partial}{\partial g^k_a} C^{\{i\}}_{\{g\}} = C^{\{i\}_k}_{\{g\}_a} \; \delta(k \in \{i\}).
\end{align}
On the right of \eqref{eq:1_der_apig_sd}, the permanent $C^{\{i\}_k}_{\{g\}_a}$ is over the set $\{i\}$ with $k$ left out, and over the set $\{g\}$ with $g_a$ left out. Further, the indicator function
\begin{align} \label{eq:gen_delta}
\delta (k \in \{i\}) = \sum_a \delta_{i_a k}
\end{align}
ensures that $k$ is in the set $\{i\}$ as otherwise the contribution should be zero. We will keep the notation $\delta (k \in \{i\})$ as it is much more clear than the right hand side of \eqref{eq:gen_delta}. The form factors are
\begin{align}
\braket{ \{h\} | S^+_k | \{g\}_a} &= \frac{\partial}{\partial g^k_a} \braket{\{h\}|\{g\}} = \sum_{\{i\}} C^{\{h\}}_{\{i\}} C^{\{i\}_k}_{\{g\}_a} \; \delta (k \in \{i\} ) \\
\braket{ \{h\} | S^+_k S^+_l | \{g\}_{a,b}} &= \frac{\partial^2}{\partial g^k_a \partial g^l_b} \braket{ \{h\} | \{g\}} = \sum_{\{i\}} C^{\{h\}}_{\{i\}} C^{\{i\}_{k,l}}_{\{g\}_{a,b} } \; \delta(k,l \in \{i\}) .
\end{align}

For the 1-DM elements we obtain
\begin{align}
\frac{1}{2} \braket{ \{h\} | \hat{n}_k | \{g\}} &= \sum_a g^k_a \sum_{\{i\}} C^{\{h\}}_{\{i\}} C^{\{i\}_k}_{\{g\}_a} \; \delta(k\in \{i\}) \\
&= \sum_{\{i\}} C^{\{h\}}_{\{i\}} \left( \sum_a g^k_a C^{\{i\}_k}_{\{g\}_a}  \right) \delta(k\in \{i\}) \label{eq:laplace_1rdm} \\
&= \sum_{\{i\}} C^{\{h\}}_{\{i\}} C^{\{i\}}_{\{g\}} \; \delta(k\in \{i\}) \label{eq:apig_ww_1rdm}
\end{align}
as the bracketed term in \eqref{eq:laplace_1rdm} is the Laplace expansion of the permanent $C^{\{i\}}_{\{g\}} $ along the $k$th column. The diagonal-correlation function is evaluated in the same manner, giving
\begin{align}
\frac{1}{4} \braket{\{h\} | \hat{n}_k \hat{n}_l | \{g\}} = \sum_{\{i\}} C^{\{h\}}_{\{i\}} C^{\{i\}}_{\{g\}} \; \delta(k,l\in \{i\}) \label{eq:apig_ww_Dij}
\end{align}
where the indicator function restricts the summation to only Slater determinants $\{i\}$ in which both $k$ and $l$ are present.

The pair-correlation function presents only minor difficulty. Using the form factor expressions, and interchanging summations, we arrive at
\begin{align} \label{eq:laplace_apig_d2p}
\braket{\{h\} | S^+_k S^-_l | \{g\}} &= \sum_{\{i\}} C^{\{h\}}_{\{i\}} \sum_a g^l_a C^{\{i\}_k}_{\{g\}_a} \; \delta (k \in \{i\}) 
- \sum_{\{i\}} C^{\{h\}}_{\{i\}} \sum_{a \neq b} g^l_a g^l_b C^{\{i\}_{k,l}}_{\{g\}_{a,b}} \; \delta(k,l\in \{i\}).
\end{align}
The first sum in \eqref{eq:laplace_apig_d2p} can be split into two sets of terms: those for which $l \in \{i\}$ and those for which $l \notin \{i\}$. The resulting sum over terms for which $l$ is a member of $\{i\}$ cancel exactly the second summation of \eqref{eq:laplace_apig_d2p} as both represent Laplace expansion of permanents in which the $l$ column appears twice. Therefore, only the single summation over terms in which $l$ is not a member of $\{i\}$ survive. The final result is not complicated, though the notation is less clear.
\begin{align} \label{eq:apig_ww_Pij}
\braket{ \{h\} | S^+_k S^-_l | \{g\}} &= \sum_{\{i\}} C^{\{h\}}_{\{i\}} C^{(\{i\} \setminus k \cup l  )}_{\{g\}} \; \delta(k\in \{i\}) \delta (l \notin \{i\})
\end{align}
The notation $\{i\} \setminus k \cup l$ is understood as the set $\{i\}$ in which $k$ is replaced by $l$. Weinhold and Wilson\cite{weinhold:1967a,weinhold:1967b} write this as a sum over two sets $\{i\}$ and $\{j\}$ such that their difference is one element:
\begin{align}
\braket{ \{h\} | S^+_k S^-_l | \{g\}} = \sum_{\{i\},\{j\} : (\{i\}\setminus k = \{j\} \setminus l  )} C^{\{h\}}_{\{i\}} C^{\{j\}}_{\{g\}}.
\end{align}
In any case, the result is not complicated. It is a sum of product of permanents which differ by one column, corresponding to the transfer of a pair from spatial orbital $l$ to spatial orbital $k$.

\subsection{Sklyanin Sum}
We now calculate the scalar product and DM elements in a different manner. Rather than projecting on Slater determinants, the APIG scalar products may be evaluated directly to yield a closed form result. The scalar product between two APIG states is
\begin{align} \label{eq:APIG_dummy}
\braket{\{h\}|\{g\}} = \sum_{i_1 \dots i_M} \sum_{j_1 \dots j_M} h^{i_1}_1 g^{j_1}_1 \dots h^{i_M}_M g^{j_M}_M \braket{\theta| S^-_{i_1}\dots S^-_{i_M} S^+_{j_M} \dots S^+_{j_1} |\theta},
\end{align}
where the summations are complete over each index. The scalar product on the right of \eqref{eq:APIG_dummy} may be evaluated ``by inspection'': if pairs of electrons were genuine bosons, it would simply be a permanent of Kronecker deltas. But pairs of electrons are not bosons and the Pauli principle prevents two pairs from occupying the same spatial orbital. The two sets of indices $\{i\}$ and $\{j\}$ must be identical, and no index may occur more than once, so the scalar product is a permanent of Kronecker deltas multiplied by a factor preventing any two pairs from occupying identical sites
\begin{align}
\braket{\theta| S^-_{i_1}\dots S^-_{i_M} S^+_{j_M} \dots S^+_{j_1} |\theta} &=
\begin{vmatrix}
\delta_{i_1 j_1} & \delta_{i_1 j_2} & \dots & \delta_{i_1 j_M} \\
\delta_{i_2 j_1} & \delta_{i_2 j_2} & \dots & \delta_{i_2 j_M} \\
\vdots & \vdots & \ddots & \vdots \\
\delta_{i_M j_1} & \delta_{i_M j_2} & \dots & \delta_{i_M j_M}
\end{vmatrix}^+
\prod_{a < b} \left( 1- \delta_{i_a i_b} \right) \\
&= \mathfrak{p}^M (\{i\},\{j\})  \mathfrak{d}^M(\{i\}), \label{eq:su2_intuitive}
\end{align}
where $\mathfrak{p}^M(\{i\},\{j\})$ represents the $M \times M$ permanent of Kronecker deltas, while $\mathfrak{d}^M(\{i\}) = \prod_{a<b} (1-\delta_{i_a i_b})$.
This result is correct, but as it is an intermediate step in our fundamental result, it will be calculated in a little more detail. We will normal order the string of objects in the scalar product and show that it produces the same result, by using the structure \eqref{eq:su2_structure}. For a string of su(2) operators, we'll define their normal order as that in which the $S^-$ are to the right (so that they destroy the vacuum), the $S^+$ are to the left (so that they destroy the dual vacuum), and the $S^z$ are in the middle (as they destroy neither the vacuum nor the dual vacuum). With this definition, the only non-vanishing contributions in the scalar product $\braket{\theta| S^-_{i_1}\dots S^-_{i_M} S^+_{j_M} \dots S^+_{j_1} |\theta}$ occur from ``full contractions'', i.e. those that contain only $S^z$. All the other terms will annihilate both vacuums.  

By normal ordering the string $S^-_{i_1}\dots S^-_{i_M} S^+_{j_M} \dots S^+_{j_1}$ there is a unique leading term of $M$ objects $2^M \mathfrak{p}^M(\{i\},\{j\})S^z_{i_1}\dots S^z_{i_M}$ that arises from moving $S^-$ past $S^+$. A factor of 2 is introduced for each exchange, and each $S^-$ must move past each $S^+$, hence the permanent $\mathfrak{p}^M(\{i\},\{j\})$.

Next, there are $\binom{M}{2}$ \emph{first descendant} terms each containing $M-1$ $S^z$ objects. These terms appear from moving an $S^z$ past an $S^{\pm}$ in addition, which causes two of the remaining $\{i\}$ indices to be identical. Evaluated directly, the numerical factor is still $2^M$, but the permanent of Kronecker deltas is damaged since there are many equivalent ways to order the indices in products of Kronecker deltas. This damage may be repaired by using e.g.
\begin{align}
2 \delta_{i_1 i_2} \delta_{i_1 j_1} \delta_{i_2 j_2} = \delta_{i_1 i_2} ( \delta_{i_1 j_1} \delta_{i_2 j_2} + \delta_{i_1 j_2} \delta_{i_2 j_1} )
\end{align}
so that $\mathfrak{p}^M(\{i\},\{j\})$ appears explicitly, giving $2^{M-1} \mathfrak{p}^M (\{i\},\{j\}) \delta_{i_1 i_2} S^z_{i_1}S^z_{i_3}\dots S^z_{i_M}$ for example. 

Second and higher descendants may be approached similarly, though the intermediate accounting becomes incredibly tedious so we will proceed directly to the result. The fully-contracted terms are in one-to-one correspondence with the $M!$ permutations of the symmetric group $\mathfrak{S}_M$. Each permutation $\sigma$ can be written as a product of $r$ disjoint $q$-cycles, $\sigma_{\alpha}$, with the multiplication understood left to right 
\begin{align}
\sigma = \sigma_1 \sigma_2 \dots \sigma_r.
\end{align}
By convention, $q$-cycles are written so that their smallest element occurs first. Each element occurs in precisely one $q$-cycle, so the sum of their lengths is M
\begin{align}
\sum^r_{\alpha=1} | \sigma_{\alpha} | = M.
\end{align}
Two operations on $q$-cycles will be necessary: define $s(\sigma_{\alpha})$ to return the smallest element of $\sigma_{\alpha}$, and $\hat{\delta} (\sigma_{\alpha})$ to return a product of Kronecker deltas of the elements $\sigma_{\alpha}$. In particular, for the $q$-cycle $(i_1 i_2 i_3 \dots i_	q)$
\begin{align}
s(i_1 i_2 i_3 \dots i_q) &= i_1 \\
\hat{\delta}(i_1 i_2 i_3 \dots i_q ) &= \delta_{i_1 i_2} \delta_{i_1 i_3} \dots \delta_{i_1 i_q}.
\end{align}
With these definitions, the fully contracted terms obtained from normal ordering are
\begin{align} \label{eq:su2_norm_order}
S^-_{i_1}\dots S^-_{i_M} S^+_{j_M} \dots S^+_{j_1} \rightarrow (-1)^M 2^M \mathfrak{p}^M (\{i\},\{j\}) \sum_{\sigma \in \mathfrak{S}_M} \prod^r_{\alpha=1} 2^{-(|\sigma_{\alpha}|-1)} \hat{\delta}(\sigma_{\alpha}) S^z_{s(\sigma_{\alpha})}.
\end{align}
Finally taking the vacuum expectation value, using the action of $S^z_i$ on the vacuum \eqref{eq:su2vac}, gives the scalar product 
\begin{align} \label{eq:su2_norm_order_scalar}
\braket{\theta | S^-_{i_1}\dots S^-_{i_M} S^+_{j_M} \dots S^+_{j_1} | \theta } =
 \mathfrak{p}^M (\{i\},\{j\}) \sum_{\sigma \in \mathfrak{S}_M} \prod^r_{\alpha=1} (-1)^{|\sigma_{\alpha}|-1} \hat{\delta}(\sigma_{\alpha}) .
\end{align}
For example, with $M=3$ the scalar product is
\begin{align}
\braket{\theta | S^-_{i_1} S^-_{i_2} S^-_{i_3} S^+_{j_3} S^+_{j_2} S^+_{j_1} | \theta } =
\mathfrak{p}^3 (\{i\},\{j\}) \left( 1
- \left( \delta_{i_1 i_2} 
+  \delta_{i_1 i_3}
+  \delta_{i_2 i_3} \right)
+ \delta_{i_1 i_2} \delta_{i_1 i_3} 
+  \delta_{i_1 i_3} \delta_{i_1 i_2}  \right).
\end{align}
To verify that the intuitive result \eqref{eq:su2_intuitive} agrees with \eqref{eq:su2_norm_order_scalar} it remains to show that
\begin{align}
\mathfrak{d}^M (\{i\}) = \sum_{\sigma \in \mathfrak{S}_M} \prod^r_{\alpha=1} (-1)^{|\sigma_{\alpha}|-1} \hat{\delta}(\sigma_{\alpha})
\end{align}
which can be proved by induction. It is obviously true for $M=2$. Assuming it to be true for $M-1$, note that from its implicit definition in \eqref{eq:su2_intuitive}, $\mathfrak{d}^M(\{i\})$ can be written
\begin{align}
\mathfrak{d}^M(\{i\}) = \mathfrak{d}^{M-1}(\{i\}) \prod^{M-1}_{a=1} ( 1 - \delta_{i_a i_M}),
\end{align}
and the product on the right is
\begin{align}
\prod^{M-1}_{a=1} ( 1 - \delta_{i_a i_M}) = 1 - \sum^{M-1}_{a=1} \delta_{i_a i_M} + \mathcal{O}(\delta^2 ).
\end{align}
The factors proportional to at least two Kronecker deltas will vanish as they imply at least two of the indices, say $i_1$ and $i_2$, from $\mathfrak{d}^{M-1}(\{i\})$ coincide, and hence when multiplied with the corresponding factor $(1-\delta_{i_1 i_2})$ will give zero since $\delta_{i_1 i_2}(1-\delta_{i_1 i_2}) = \delta_{i_1 i_2} - \delta^2_{i_1 i_2} =0$. Finally, the factor 1 returns the original $(M-1)!$ permutations (of $\mathfrak{S}_{M-1}$) while the sum of Kronecker deltas multiplied by $\mathfrak{d}^{M-1}(\{i\})$ yields $(M-1)\cdot (M-1)!$ new terms, yielding in total the $M!$ permutations (of $\mathfrak{S}_M$) required, with their correct signs. Thus the intuitive result \eqref{eq:su2_intuitive} and \eqref{eq:su2_norm_order_scalar} are equivalent. The advantage of \eqref{eq:su2_norm_order} is that it relies only on the structure of the Lie algebra \eqref{eq:su2_structure} and \emph{not} the representations \eqref{eq:su2_objects}. Thus, it is also applicable to other representations of su(2).

The scalar product \eqref{eq:APIG_dummy} can now be simplified. Again, if pairs of electrons were genuine bosons, then $\mathfrak{d}^M(\{i\})$ would be identically one, and the scalar product $\braket{\{h\}|\{g\}}$ would be a permanent of contributions
\begin{align}
\Gamma (h_a,g_b) = \sum_i h^i_a g^i_b,
\end{align}
which we call  rank-1 contractions, corresponding to the elements of $\mathfrak{p}^M(\{i\},\{j\})$. Each permutation in $\mathfrak{d}^M(\{i\})$ contributes an additional term, which can be factored into contributions from its constituent disjoint $q$-cycles. The contributions are not unique as each $q$-cycle contribution corresponds to $q$ indices among $\{i\}$ being identical. Each occurs $(q-1)!$ times: the smallest element $s(\sigma_{\alpha})$ is fixed while the other $q-1$ indices may be permuted to yield distinct $q$-cycles but equivalent contributions, all weighted by $(-1)^{q-1}$. When multiplied with $\mathfrak{p}^M(\{i\},\{j\})$, a sum over the symmetric group, the permutations which permute only indices of these particular $q$-cycles leave the contribution invariant, each occurring $q!$ times. Define the rank-$q$ contraction as
\begin{align} \label{eq:apig_gamma}
\Gamma \left( h_{a_1}, \dots h_{a_q} , g_{b_1}, \dots , g_{b_q}\right)  &= (-1)^{(q-1)} q! (q-1)!\sum_i h^i_{a_1} \dots h^i_{a_q} g^i_{b_1}\dots g^i_{b_q} \\
& = \sum_i \Gamma^i \left( h_{a_1}, \dots h_{a_q} , g_{b_1}, \dots , g_{b_q}\right) \label{eq:gamma_short}
\end{align}
where in the second line the shorthand $\Gamma^i$ was introduced for later use. The final result for the scalar product is
\begin{align} \label{eq:apig_sklyanin}
\braket{\{h\} | \{g\}} = \sum_{\mathcal{P}} \prod_{P \in \mathcal{P}}  \Gamma\left( \{h\}_P \cup \{g\}_P \right)
\end{align}
which we refer to as the Sklyanin sum, as Sklyanin obtained the corresponding case for off-shell RG states,\cite{sklyanin:1999} and we henceforth adopt his notation. The sets of geminal coefficients $\{h\}$ and $\{g\}$ are each split by $\mathcal{P}$ into disjoint partitions $P$. $\mathcal{P}$ is the set of all such collections of partitions such that:
\begin{align}
|\{h \}_P| &= | \{g\}_P|, \quad \forall P \in \mathcal{P} \\
\bigcup_{P \in \mathcal{P}} \{h \}_P = \{ h \}, &\quad
\bigcup_{P \in \mathcal{P}} \{g \}_P = \{ g \} \\
\sum_{P \in \mathcal{P}} | \{h\}_P| = |\{h\}| &= \sum_{P \in \mathcal{P}} | \{g\}_P|  = |\{g\}| = M
\end{align}
and for distinct $P$ and $P'$ the partitions are necessarily disjoint
\begin{align}
\{h\}_P \cap \{h\}_{P'} = \emptyset, \quad \{g \}_P \cap \{g \}_{P'} = \emptyset.
\end{align}
As this notation will very quickly get out of control, we will abbreviate the contractions as
\begin{align}
\Gamma \left( \{h\}_P \cup \{g\}_P \right) \rightarrow \Gamma_P,
\end{align}
it being understood that the partition $P$ corresponds to the elements $\{h\}_P \cup \{g\}_P$. For $M=2$, the scalar product is
\begin{align}
\braket{\{h\} | \{g\}} = \Gamma(h_1, g_1)\Gamma (h_2,g_2) + \Gamma(h_1,g_2) \Gamma(h_2,g_1) + \Gamma(h_1,h_2,g_1,g_2)
\end{align}
for example.

The expression \eqref{eq:apig_sklyanin} is the analogue of Wick's theorem for fundamental representations of su(2): pairs of fermions \eqref{eq:su2_objects}, or $\frac{1}{2}$-spins \eqref{eq:su2_objects_spin}. The single-particle basis information enters only in complete summations (as for Slavnov's theorem) and the physical picture of pairs is not immediately lost by projecting onto Slater determinants. Equation \eqref{eq:apig_sklyanin} has been verified numerically against the Slater determinant result \eqref{eq:apig_slavnov_sd}.

To ensure a complete list of rank-$q$ contractions, it is convenient to arrange them in a matrix of size $\binom{M}{q}\times \binom{M}{q}$, e.g. for $q=2$
\begin{align}
\begin{pmatrix}
\Gamma (h_1,h_2,g_1,g_2) & \Gamma (h_1,h_2,g_1,g_3) & \dots & \Gamma (h_1,h_2,g_{M-1},g_M) \\
\Gamma (h_1,h_3,g_1,g_2) & \Gamma (h_1,h_3,g_1,g_3) & \dots & \Gamma (h_1,h_3,g_{M-1},g_M) \\
\vdots & \vdots & \ddots &\vdots\\
\Gamma (h_{M-1},h_M,g_1,g_2) & \Gamma (h_{M-1},h_M,g_1,g_3) & \dots & \Gamma (h_{M-1},h_M,g_{M-1},g_M)
\end{pmatrix},
\end{align}
where each row represents a particular choice of 2 $h$'s and each column represents a choice of 2 $g$'s. The structure for higher ranks is analogous, with each row a particular choice of $q$ $h$'s and each column a choice of $q$ $g$'s. It is natural to refer to the diagonal elements of these matrices as the \emph{diagonal} rank-$q$ contractions. These matrices could of course be diagonalized to simplify the Sklyanin sum, though this is intractable as their ranks grow like squares of binomial coefficients. We will see in section \ref{sec:APSG} that strong-orthogonality implies that only the diagonal rank-1 contractions survive. When $\{h\} = \{g\}$, the matrices of contractions will obviously become symmetric.

To evaluate form factors, we must differentiate \eqref{eq:apig_sklyanin} with respect to geminal coefficients $g^k_a$. In each summand of \eqref{eq:apig_sklyanin}, the coefficient $g^k_a$ occurs exactly once, so that the first derivatives have a clean expression:
\begin{align}
 \frac{\partial}{\partial g^k_a} \braket{\{h\}|\{g\}} = \sum_{\mathcal{P}} \frac{\partial \Gamma_{P(g_a)}}{\partial g^k_a} \prod_{P (\neq P(g_a)) \in \mathcal{P}} \Gamma_P 
\end{align}
where $P(g_a)$ is the element of $\mathcal{P}$ containing the coefficients $g_a$. 

First and second derivatives of the scalar product \eqref{eq:apig_sklyanin} only depend on first derivatives of \eqref{eq:apig_gamma}, which is easily evaluated:
\begin{align}
\frac{\partial \Gamma_P }{\partial g^k_c} = (-1)^{(\vert \{h\}_P \vert -1)} \vert \{h\}_P\vert ! ( \vert \{h\}_P \vert -1 )!   \prod_{\substack{a \in \{h\}_P \\ b (\neq c) \in \{g\}_P} } h^k_a g^k_b
=  \frac{1}{g^k_c} \Gamma^k_P,
\end{align}
where the functions $\Gamma^k_P$ were implicitly defined above in \eqref{eq:gamma_short}. Second derivatives are simple as well. Since we are a priori only considering $k\neq l$, only collections partitions for which $g^k_a$ and $g^l_b$ occur in \emph{separate} elements will give non-zero contributions. Thus with a summation over only such collections of partitions $\mathcal{P}'$,
\begin{align}
 \frac{\partial^2}{\partial g^k_a \partial g^l_b}  \braket{\{h\}|\{g\}} &= \sum_{\mathcal{P}'} \frac{\partial \Gamma_{P(g_a)}}{\partial g^k_a} \frac{\partial \Gamma_{P(g_b)} }{\partial g^l_b} \prod_{P (\neq P(g_a),P(g_b))\in \mathcal{P}'} \Gamma_P \\
&= \frac{1}{g^k_a g^l_b} \sum_{\mathcal{P}'}  \Gamma^k_{P(g_a)}  \Gamma^l_{P(g_b)}  \prod_{P (\neq P(g_a),P(g_b))\in \mathcal{P}'} \Gamma_P .
\end{align}
Evaluating summations is now trivial:
\begin{align}
\frac{1}{2}\braket{\{h\}| \hat{n}_k | \{g\}} &= \sum_a \sum_{\mathcal{P}} \Gamma^k_{P(g_a)} \prod_{P (\neq P(g_a)) \in \mathcal{P}} \Gamma_P \label{eq:apig_sklyanin_1rdm}\\
\frac{1}{4} \braket{\{h\}|\hat{n}_k \hat{n}_l | \{g\}} &= \sum_{a\neq b} \sum_{\mathcal{P}'} \Gamma^k_{P(g_a)} \Gamma^l_{P(g_b)} \prod_{P (\neq P(g_a),P(g_b))\in \mathcal{P}'} \label{eq:apig_sklyanin_Dij} \Gamma_P \\
\braket{\{h\} |S^+_k S^-_l | \{g\}} &= \sum_a \frac{g^l_a}{g^k_a} \sum_{\mathcal{P}} \Gamma^k_{P(g_a)} \prod_{P (\neq P(g_a)) \in \mathcal{P}} \Gamma_P \nonumber \\
&- \sum_{a\neq b} \frac{g^l_a}{g^k_a} \sum_{\mathcal{P}'} \Gamma^k_{P(g_a)} \Gamma^l_{P(g_b)} \prod_{P (\neq P(g_a),P(g_b))\in \mathcal{P}'} \Gamma_P  \label{eq:better_APIG_Pkl}
\end{align}
Equations \eqref{eq:apig_sklyanin_1rdm}, \eqref{eq:apig_sklyanin_Dij} and \eqref{eq:better_APIG_Pkl} have been verified numerically against \eqref{eq:apig_ww_1rdm}, \eqref{eq:apig_ww_Dij} and \eqref{eq:apig_ww_Pij}. These expressions correspond directly to the on-shell RG results. Scalar products and correlation functions are computable from the primitives $\Gamma^k$. Setting $\{h\}\rightarrow\{g\}$ presents no further simplification or difficulty, so we finish with these expressions as they represent both RDM elements if $\{h\}=\{g\}$ and TDM elements otherwise.

\subsection{Alternative expression for the pair-correlation function}
The form factor approach reduces the expression for $\braket{\{h\}|S^+_k S^-_l | \{g\}}$ to two summations over $M$-pair scalar products. This necessarily privileges one of the sets over the other, and the result appears asymmetric in $k$ and $l$ even though it isn't. Instead, we could act to the right with $S^-_l$ and to the left with $S^+_k$. The result will be summations over $(M-1)$-pair scalar products which are obviously symmetric in $k$ and $l$. This approach was not taken for on-shell states as one of the sets of geminal coefficients \emph{is} special: they must satisfy Richardson's equations. Acting with $S^-_l$ to the right and $S^+_k$ to the left gives
\begin{align}
S^-_l \ket{\{g\}} &= \sum_a g^l_a \left( 1 - \sum_{b (\neq a)} g^l_b \frac{\partial}{\partial g^l_b} \right) \ket{\{g\}_a} \\
\bra{\{h\} } S^+_k &= \sum_c h^k_c \left( 1 - \sum_{d (\neq c)} h^k_d \frac{\partial}{\partial h^k_d} \right) \bra{\{h\}_c}.
\end{align}
The $(M-1)$-pair scalar products $\braket{\{h\}_c|\{g\}_a}$ are evaluated as Sklyanin sums in which the sets of partitions $\mathcal{P}_{\bar{g}_a\bar{h}_c}$ leave out the coefficients $h_c$ and $g_a$. The final expression is much less compact, but once again the principal difficulty is the notation. The result itself 
\begin{align}
\braket	{ \{h\} | S^+_k S^-_l | \{g\}} &= \sum_{ac} h^k_c g^l_a \sum_{\mathcal{P}_{\bar{g}_a\bar{h}_c} } \prod_{P \in \mathcal{P}_{\bar{g}_a\bar{h}_c}} \Gamma_P \nonumber \\
&- \sum_{ac}  h^k_c g^l_a \sum_{d (\neq c)} \sum_{\mathcal{P}_{\bar{g}_a\bar{h}_c} } \Gamma^k_{P(h_d)} \prod_{P (\neq P(h_d)) \in \mathcal{P}_{\bar{g}_a\bar{h}_c}} \Gamma_P  \nonumber \\
&- \sum_{ac} h^k_c g^l_a \sum_{b (\neq a)}  \sum_{\mathcal{P}_{\bar{g}_a\bar{h}_c} } \Gamma^l_{P(g_b)} \prod_{P (\neq P(g_b)) \in \mathcal{P}_{\bar{g}_a\bar{h}_c}} \Gamma_P \nonumber \\
&+ \sum_{ac} h^k_c g^l_a \sum_{b (\neq a)} \sum_{d (\neq c)}   \sum_{\mathcal{P}^{'}_{\bar{g}_a\bar{h}_c} } \Gamma^k_{P(h_d)} \Gamma^l_{P(g_b)}  \prod_{P (\neq P(h_d),P(g_b)) \in \mathcal{P}^{'}_{\bar{g}_a\bar{h}_c}} \Gamma_P
\end{align}
is not complicated. As in the previous expressions, the final line involves summation only over sets of partitions $\mathcal{P}^{'}_{\bar{g}_a\bar{h}_c}$ in which $h^k_d$ and $g^l_b$ occur in separate partitions. While this expression is clearly symmetric, the expression \eqref{eq:better_APIG_Pkl} is easier to evaluate. We will therefore not consider this approach any further.

In the following sections we will outline three degenerate cases for which the Sklyanin sums reduce to feasible expressions. This is accomplished in three different ways. For RG states, the rank of the contractions reduce to one. (This alone does not yield a tractable sum, the RG states must be on-shell.) For AGP, the contractions lead to a simple recursion. For APSG, the non-zero contractions are sparse.

\section{Off-Shell RG States: Rank-reduction} \label{sec:offshell}
On-shell RG states are a specific limit of the ABA solution to the 6-vertex model.\cite{yang:1966a,yang:1966b,yang:1966c,lieb:1967a,lieb:1967b,lieb:1967c,takhtadzhan:1979,faddeev:1981} In a landmark contribution, Korepin calculated the scalar products and norms of off-shell ABA states.\cite{korepin:1982} It was shown that the scalar products depended upon $4M$ numbers: the rapidities and vacuum eigenvalues of the two states. Slavnov showed that when one of the states was on-shell, the scalar product depended on $3M$ numbers (since the on-shell conditions force a dependence between the rapidities and vacuum eigenvalues), and could be computed as a single determinant.\cite{slavnov:1989} The case of Slavnov's theorem \eqref{eq:slavnov} is the so-called ``quasi-classical'' limit first written explicitly in ref.\cite{zhou:2002} 

The key property of RG states is the Gaudin algebra structure: for any complex number $u$, define three objects $S^+(u),S^-(u),S^z(u)$ with commutators
\begin{subequations}\label{eq:XXZ_gaudin}
\begin{align}
[S^+(u), S^-(v)] &= 2 Z(u,v) (S^z(u) - S^z(v)) \\
[S^z(u), S^{\pm}(v)] &= \pm X(u,v) (S^{\pm}(u) - S^{\pm}(v))
\end{align}
\end{subequations}
in terms of some arbitrary functions $X(u,v)$ and $Z(u,v)$. For these objects to close a Lie algebra, Jacobi identities must be satisfied by double commutators to ensure that the action is associative, which forces 
\begin{align} \label{eq:XXZ_cond}
X(u,v) X(v,w) = X(u,w) \left( Z(u,v) + Z(v,w) \right).
\end{align}
These conditions were originally obtained by Gaudin as integrability conditions for a collection of spin Hamiltonians.\cite{gaudin:1976} The simplest solution is isotropic
\begin{align}
X(u,v) = Z(u,v) = \frac{1}{u-v}
\end{align}
in terms of rational functions, while there are also anisotropic solutions 
\begin{align}
X(u,v) = \frac{1}{\text{sin[h]}(u-v)}, \quad Z(u,v) = \text{cot[h]}(u-v)
\end{align}
in terms of trigonometric or [hyperbolic] functions. Other parametrizations are possible, though all reduce to one of these cases.\cite{dukelsky:2004,ortiz:2005,ibanez:2009,dunning:2010,dukelsky:2011,vanraemdonck:2014,claeys:2015,claeys:2017a} The isotropic case of \eqref{eq:XXZ_cond} is precisely the partial fraction decomposition. For APIG generally, such a structure does not exist: the loosest conditions possible leading to a Lie algebra linear in the geminal coefficients \emph{are} the XXZ conditions \eqref{eq:XXZ_cond}. A representation of the objects \eqref{eq:XXZ_gaudin} in terms of the pair operators can be chosen
\begin{align}
S^{\pm}(u) &= \sum_i X(u,\varepsilon_i) S^{\pm}_i \\
S^z (u) &= \frac{1}{g} - \sum_i Z(u,\varepsilon_i) S^z_i
\end{align}
in terms of a set of arbitrary real numbers $\{\varepsilon\}$ and a constant $g$. For the remainder of this contribution we will only treat the rational case. No substantial complication arises for the anisotropic cases, though the results become substantially more opaque. The fundamental simplification for the Sklyanin sum occurs due to partial fractions (in eq. \eqref{eq:partial_frac_simp}), which is the isotropic case of \eqref{eq:XXZ_cond}.

\subsection{Slater determinant projection}
The results in terms of Slater determinants are straightforward limits of the APIG expressions, so we will proceed directly to the results. In a basis of Slater determinants, the states $\ket{\{u\}}$ are
\begin{align}
\ket{\{u\}} =  \sum_{ \{i\} } C^{\{u\}}_{\{i\}} \ket{\{i\} }
\end{align}
where the expansion coefficients $C^{\{u\}}_{\{i\}} \equiv \braket{\{u\} | \{i\} } $ are $M\times M$ Cauchy permanents:
\begin{align}
C^{\{u\}}_{\{i\}} = \underset{i,a} {\text{per}} \left( \frac{1}{u_a - \varepsilon_i} \right)
 = \sum_{\sigma \in \mathfrak{S}_M} \prod^M_{a=1} \frac{1}{u_a - \varepsilon_{\sigma(i_a)}} = 
\begin{vmatrix}
\frac{1}{u_1 - \varepsilon_{i_1}} & \frac{1}{u_1 - \varepsilon_{i_2}} & \dots & \frac{1}{u_1 - \varepsilon_{i_M}} \\
\frac{1}{u_2 - \varepsilon_{i_1}} & \frac{1}{u_2 - \varepsilon_{i_2}} & \dots & \frac{1}{u_2 - \varepsilon_{i_M}} \\
\vdots & \vdots & \ddots & \vdots \\
\frac{1}{u_M - \varepsilon_{i_1}} & \frac{1}{u_M - \varepsilon_{i_2}} & \dots & \frac{1}{u_M  - \varepsilon_{i_M}}
\end{vmatrix}^+.
\end{align}
The summation in $\sigma$ is again over the entire symmetric group $\mathfrak{S}_M$. Individual Cauchy permanents are computable with Borchardt's theorem,\cite{borchardt:1857}
\begin{align}
\det_{i,a} \left( \frac{1}{u_a - \varepsilon_i} \right) \underset{i,a} {\text{per}} \left( \frac{1}{u_a - \varepsilon_i} \right) = \det_{i,a} \left( \frac{1}{(u_a - \varepsilon_i)^2} \right)
\end{align}
or more simply\cite{faribault:2012,gaudin_book,claeys:2017b}
\begin{align} \label{eq:far_det}
\underset{i,a} {\text{per}} \left( \frac{1}{u_a - \varepsilon_i} \right) = \det F
\end{align}
with
\begin{align}
F_{ij} =
\begin{cases}
\sum_a \frac{1}{\varepsilon_i - u_a} - \sum_{k \neq i} \frac{1}{\varepsilon_i - \varepsilon_k}, \quad & i=j \\
- \frac{1}{\varepsilon_i - \varepsilon_j}, \quad & i\neq j
\end{cases}.
\end{align}
Individual Cauchy permanents are feasible to evaluate, but there remain $\binom{N}{M}$ to compute so scalar products between off-shell RG vectors remain intractable. 

The scalar products and DM elements are the same as for APIG
\begin{align}
\braket{ \{v\} | \{u\}} &= \sum_{\{i\}} C^{\{i\}}_{\{v\}} C^{\{u\}}_{\{i\}} \label{eq:off_slater_scalar} \\
\frac{1}{2} \braket{ \{v\} | \hat{n}_k | \{u\}} &= \sum_{\{i\}} C^{\{v\}}_{\{i\}} C^{\{i\}}_{\{u\}} \; \delta(k\in \{i\}) \label{eq:off_ww_1rdm}\\
\frac{1}{4} \braket{ \{v\} | \hat{n}_k \hat{n}_l | \{u\}} &= \sum_{\{i\}} C^{\{v\}}_{\{i\}} C^{\{i\}}_{\{u\}} \; \delta(k,l \in \{i\}) \label{eq:off_ww_Dij} \\
\braket{\{v\} | S^+_k S^-_l | \{u\}} &= \sum_{\{i\}} C^{\{v\}}_{\{i\}} C^{(\{i\} \setminus k \cup l  )}_{\{u\}} \; \delta(k\in \{i\}) \delta (l \notin \{i\}) \label{eq:off_ww_Pij}.
\end{align}
In the Slater determinant basis there is again no further simplification, or complication, in the limit $\{v\} = \{u\}$.

\subsection{Sklyanin Sum}
Sklyanin obtained a closed-form expression for the scalar product using a generating function approach.\cite{sklyanin:1999} The stated purpose of that contribution was to develop the off-shell scalar products for RG states in the same manner as Korepin's combinatorial approach for the 6-vertex ABA states. Sklyanin's result is of exactly the same shape as our APIG scalar product. We arrived at \eqref{eq:apig_sklyanin} independently, and seeing the similarity was immediately indicative that our result was correct. Sklyanin's result differs only in the form of the contractions, which simplify due to the rational structure of the geminal coefficients. Specifically, the individual summands in \eqref{eq:gamma_short} become
\begin{align}
\Gamma^i_P = \prod_{\substack{ a \in \{v\}_P \\ b \in \{u\}_P } } \frac{1}{(v_a - \varepsilon_i)(u_b - \varepsilon_i)}.
\end{align}
Denote by the union of the two sets $\{u\}_P \cup \{v\}_P = \mathcal{W}_P$ so that
\begin{align} \label{eq:partial_frac_simp}
\Gamma^i_P = \prod_{w \in \mathcal{W}_P} \frac{1}{(w - \varepsilon_i)} = \sum_{w\in \mathcal{W}_P} \frac{A_w}{(w- \varepsilon_i)}
\end{align}
where in the last equality the product has been separated by partial fractions with coefficients $A_w$. These coefficients may be found directly. Adding the fractions gives
\begin{align}
1 = \sum_{w \in \mathcal{W}_P} A_w \prod_{w' \neq w} (w' - \varepsilon_i) \label{eq:partial_fractions}
\end{align}
and since this must hold for each value of $\varepsilon_i$ we can choose it to be each of the $w$ individually, with each choice leaving the single non-zero term
\begin{align}
A_w = \frac{1}{\prod_{w' \neq w} (w'-w)} = \frac{(-1)^{(|\{v\}_P | -1) + |\{v\}_P| }}{\prod_{w'\neq w} (w-w')}.
\end{align}
Taking the sum gives
\begin{align}
\sum_i \Gamma^i_P = \sum_{w \in \mathcal{W}_P} A_w \left( \frac{2}{g} + \sum_i \frac{1}{(w-\varepsilon_i)} \right) = 2 \sum_{w \in \mathcal{W}_P} A_w  \alpha (w)
\end{align}
where the factor $\frac{2}{g}$ has been included so that the bracketed term is twice the vacuum eigenvalue
\begin{align}
S^z(w) \ket{\theta} = \alpha (w) \ket{\theta}.
\end{align}
This may be done since expanding \eqref{eq:partial_fractions}, the term proportional to the highest power of $\varepsilon_i$ is $\sum_{w \in \mathcal{W}_P} A_w$ and must therefore vanish. Rather than $\Gamma_P$ it is convenient to use the $\Lambda$-contraction
\begin{align}
\Lambda (\mathcal{W}_P) = \sum_{w \in \mathcal{W}_P} \frac{2 \alpha(w)}{\prod_{w' \neq w} (w-w')}.
\end{align} 
The definition of $\Gamma_P$ involves the factor $(-1)^{|\{v\}_P| - 1}$ which when combined with the factor in $A_w$ gives the factor $(-1)^{|\{v\}_P|}$. In the resulting Sklyanin sum, these factors can be combined since $\sum_{P \in \mathcal{P}} |\{v\}_P| = M$, and putting these results together gives the scalar product
\begin{align}
\braket{ \{ v \} | \{ u \} } = \left( -1 \right)^M \sum_{\mathcal{P}} \prod_{P \in \mathcal{P}} | \{ v \}_P |! \left( | \{ v \}_P |-1 \right)! \Lambda \left( \{ u \}_P \cup \{ v \}_P \right)
\end{align}
which is precisely Sklyanin's result.

The key property of the $\Lambda$-contraction is that \emph{all higher-rank contractions reduce to rank-1 contractions}
\begin{align}
\Lambda ( \{u\}_P \cup \{v\}_P ) = \sum_{\substack{ u\in \{u\}_P \\ v \in \{v\}_P } } \frac{\Lambda (u,v)}{\prod_{u' (\neq u) \in \{u\}_P } (u-u') \prod_{v' (\neq v) \in \{v\}_P} (v-v') },
\end{align}
where the rank-one contractions are
\begin{align}
\Lambda(u,v) = 2 \frac{\alpha(u) - \alpha(v)}{u-v}.
\end{align}
It is therefore possible to reduce the Sklyanin sum to a cleaner expression in terms of only rank-1 contractions. Different final expressions are possible, though all remain intractable computationally, so we will only mention Richardson's original result\cite{richardson:1965}
\begin{align} \label{eq:rich_det_sum}
\braket{\{v\}| \{u\}} = (-1)^M \sum_{\sigma \in \mathfrak{S}_M} \det R^{\sigma} 
\end{align}
where the elements of the matrices $R^{\sigma}$ are:
\begin{align}
R^{\sigma}_{ab} = 
\begin{cases}
 \Lambda(u_a,v_{\sigma(a)})  + \sum_{c \neq a} \frac{2}{(u_a - u_c)(v_{\sigma(a)} - v_{\sigma(c)}  )} ,&\quad a=b \\
-\frac{2}{(u_a - u_b) (v_{\sigma(a)} - v_{\sigma(b)}  )} ,& \quad a\neq b
\end{cases}.
\end{align}
This is the cleanest expression for the overlap between off-shell rational RG states, and we will use it to compute the DM elements. Equation \eqref{eq:rich_det_sum} has been verified numerically against \eqref{eq:off_slater_scalar}.

To compute the norm $\braket{\{v\}|\{v\}}$, all that is required is to replace $\{u\}\rightarrow\{v\}$ in the sum \eqref{eq:rich_det_sum} keeping in mind that
\begin{align} \label{eq:lam_diag}
\Lambda (v_a, v_b) = \begin{cases} 2 \frac{\partial \alpha (v_a)}{\partial v_a}, &\quad a = b \\
2 \frac{\alpha(v_a) - \alpha(v_b)}{(v_a - v_b)}, &\quad a \neq b,
\end{cases}
\end{align}
where in this particular case
\begin{align}
\frac{\partial \alpha (v)}{\partial v} = - \sum_i \frac{1}{(v - \varepsilon_i)^2}.
\end{align}
Just as for the on-shell case, the form factors are evaluated as the residues of the scalar product at the simple poles
\begin{align}
\braket{\{v\} | S^+_k | \{u\}_a} &= \lim_{u_a \rightarrow \varepsilon_k} (u_a - \varepsilon_k) \braket{ \{v\} | \{u\}} \\
\braket{\{v\} | S^+_k S^+_l | \{u\}_{a,b}} &= \lim_{u_a \rightarrow \varepsilon_k} \lim_{u_b \rightarrow \varepsilon_l} (u_a - \varepsilon_k) (u_b - \varepsilon_l) \braket{ \{v\} | \{u\}},
\end{align}
which are obtained from the determinants in \eqref{eq:rich_det_sum}. For the rapidity $u_a$, each determinant $R^{\sigma}$ has a simple pole in the $a$th diagonal element, with residue given by an $(M-1)\times (M-1)$ determinant. Specifically,
\begin{align}
\braket{\{v\} | S^+_k | \{u\}_a} = (-1)^M \sum_{\sigma \in \mathfrak{S}_M} \frac{1}{\varepsilon_k - v_{\sigma(a)}} \det R^{\sigma,ak}
\end{align}
where the matrix $R^{\sigma,ak}$ is the matrix $R^{\sigma}$ with the $a$th row and the $a$th column removed, and the diagonal elements are modified to be
\begin{align}
R^{\sigma,ak}_{cc} = \Lambda(u_c , v_{\sigma(c)}) + \sum_{c' (\neq c,a)} \frac{2}{(u_c - u_{c'})(v_{\sigma(c)} - v_{\sigma(c')})} + \frac{2}{(u_c - \varepsilon_k)(v_{\sigma(c)} - v_{\sigma(a)} )}.
\end{align}
Likewise,
\begin{align}
\braket{\{v\} | S^+_k S^+_l | \{u\}_{a,b}} = (-1)^M \sum_{\sigma \in \mathfrak{S}_M} \frac{1}{(\varepsilon_k - v_{\sigma(a)})(\varepsilon_l - v_{\sigma(b)})} \det R^{\sigma,akbl}
\end{align}
where $R^{\sigma,akbl}$ is $R^{\sigma}$ without the $a$th and $b$th rows and columns, and diagonal elements
\begin{align}
R^{\sigma,akbl}_{cc} &= \Lambda(u_c , v_{\sigma(c)}) + \sum_{c' (\neq c,a,b)} \frac{2}{(u_c - u_{c'})(v_{\sigma(c)} - v_{\sigma(c')})}  \nonumber \\
&+ \frac{2}{(u_c - \varepsilon_k)(v_{\sigma(c)} - v_{\sigma(a)} )} + \frac{2}{(u_c - \varepsilon_l)(v_{\sigma(c)} - v_{\sigma(b)} )}.
\end{align}

Finally, the density matrix elements are:
\begin{align}
\frac{1}{2}\braket{\{v\} | \hat{n}_k | \{u\}} &= (-1)^M \sum_a \sum_{\sigma \in \mathfrak{S}_M} \frac{\det R^{\sigma,ak}}{(u_a - \varepsilon_k)(\varepsilon_k - v_{\sigma(a)})} \label{eq:off_sklyanin_1rdm} \\
\frac{1}{4}\braket{ \{v\}| \hat{n}_k \hat{n}_l | \{u\}} &= (-1)^M \sum_{a \neq b} \sum_{\sigma \in \mathfrak{S}_M} \frac{\det R^{\sigma,akbl}}{(u_a - \varepsilon_k)(\varepsilon_k - v_{\sigma(a)})(u_b - \varepsilon_l)(\varepsilon_l - v_{\sigma(b)})} \label{eq:off_sklyanin_Dij} \\
\braket{\{v\} | S^+_k S^-_l | \{u\}} &= (-1)^M \sum_a \sum_{\sigma \in \mathfrak{S}_M} \frac{\det R^{\sigma,ak}}{(u_a - \varepsilon_l)(\varepsilon_k - v_{\sigma(a)})} \nonumber \\
&- (-1)^M \sum_{a \neq b} \sum_{\sigma \in \mathfrak{S}_M} \frac{\det R^{\sigma,akbl}}{(u_a-\varepsilon_l)(\varepsilon_k - v_{\sigma(a)})(u_b - \varepsilon_l)(\varepsilon_l - v_{\sigma(b)})} \label{eq:off_sklyanin_Pij}.
\end{align}
Equations \eqref{eq:off_sklyanin_1rdm}, \eqref{eq:off_sklyanin_Dij} and \eqref{eq:off_sklyanin_Pij} have been numerically verified against \eqref{eq:off_ww_1rdm}, \eqref{eq:off_ww_Dij} and \eqref{eq:off_ww_Pij}. To compute RDM elements, the rank-1 contractions become \eqref{eq:lam_diag}.

\subsection{Reduction to Slavnov determinant}
The Sklyanin sum must become the Slavnov result when one of the sets of rapidities are on-shell. We will sketch how this happens as the details are rather complicated. In the Sklyanin sum, the only terms that will matter are the leading ones, specifically the permanent of rank-one contractions $\underset{ab}{\text{per}}\; \Lambda(u_b,v_a)$. All other terms are unwanted. They are of course present, and non-zero, but we will separate them for now. When the set $\{v\} \rightarrow \{\tilde{v}\}$ are on-shell, i.e. they satisfy Richardson's equations
\begin{align}
\alpha (\tilde{v}_a ) = \sum_{c (\neq a)} \frac{1}{\tilde{v}_a - \tilde{v}_c},
\end{align}
the rank-1 contractions become
\begin{align}
\Lambda (u_b, \tilde{v}_a) = \frac{2 \alpha(u_b)}{u_b - \tilde{v}_a} - \frac{1}{u_b - \tilde{v}_a} \sum_{c (\neq a)} \frac{2}{u_b - \tilde{v}_c} - \sum_{c (\neq a)} \frac{2}{(u_b - \tilde{v}_c)(\tilde{v}_a-\tilde{v}_c)}.
\end{align}
The first two terms are of interest, while the third term is not. Thus, we can separate all terms proportional to the third term and collect them with the rest of the unwanted terms. The Sklyanin sum can then be written 
\begin{align}
\braket{ \{ \tilde{v}\} | \{u\} } = \text{per}(L) + \phi_S
\end{align}
with
\begin{align}
L_{ab} = \frac{1}{(\tilde{v}_a - u_b)} \left( 2 \alpha(u_b) - \sum_{c (\neq a)} \frac{2}{(u_b - \tilde{v}_c )} \right)
\end{align}
and $\phi_S$ is all the other terms. The factor $(-1)^M$ has been absorbed by switching the order of the denominators $(\tilde{v}_a - u_b)$. 

Muir's theorem\cite{muir:1897} for two matrices $A$ and $B$
\begin{align}
\det (A)\; \text{per} (B) = \sum_{\sigma \in \mathfrak{S}_M} \det (A * B_{\sigma})
\end{align}
is a summation over the $M!$ elements of $\mathfrak{S}_M$. The matrix $B_{\sigma}$ is $B$ whose columns have been permuted by $\sigma$ and $*$ denotes the Hadamard (element-wise) matrix product. If we choose the matrix $A$ to be the Cauchy matrix
\begin{align}
A_{ab} = \frac{1}{\tilde{v}_a - u_b}
\end{align}
then its determinant is known in closed form
\begin{align}
\det A = \frac{\prod_{a<b} (u_a - u_b) (\tilde{v}_b - \tilde{v}_a)}{\prod_{ab} (\tilde{v}_a - u_b)}.
\end{align}
Now, Muir's theorem gives
\begin{align}
\text{per}  (L) &= \frac{1}{\det (A)} \sum_{\sigma \in \mathfrak{S}_M} \det (A * L_{\sigma}) \\
&= \frac{\det( A * L)}{\det(A)} + \phi_M
\end{align}
where in the second line we have separated the identity permutation, and grouped all the rest of the terms into $\phi_M$. Notice that the first term is precisely the Slavnov result \eqref{eq:slavnov}:
\begin{align}
\frac{\det (A * L)}{\det (A)} = K \det (J)
\end{align}
and so
\begin{align}
\braket{\{\tilde{v}\} | \{u\}} = K \det (J) + \phi_M + \phi_S.
\end{align}
We can conclude that the unwanted terms from the Sklyanin sum, $\phi_S$, and the unwanted terms from Muir's theorem, $\phi_M$, cancel one another. If Slavnov's theorem were unknown it might be productive to show in detail how this cancellation occurs. Slavnov's theorem can be derived much more directly in other manners.\cite{belliard:2019}

\section{AGP: Recursion} \label{sec:agp}
AGP is the limiting case of APIG, when all the geminals are identical. It is possible to define AGPs for more general types of pairs, but this may always be reduced to closed-shell pair creators by an orbital transformation. Our choice of notation suggests the labelling
\begin{align}
\ket{g^M} = \left( \sum_i g^i S^+_i \right)^M \ket{\theta}.
\end{align}

AGP is a wavefunction ansatz with a long history in quantum chemistry.\cite{coleman:1965,ortiz:1981,sarma:1989,coleman:1997} Coleman in particular was quite interested as its 2-RDM has a macroscopic eigenvalue, which permits off-diagonal-long-range order (ODLRO), a necessary property for superconductivity.\cite{yang:1962,coleman:1989,dunne:1994} This should not be surprising as AGP is the result of projecting a particle number upon the BCS ansatz.\cite{schrieffer_book,dukelsky:2016} However, AGP is not size-consistent,\cite{linderberg:1980} which is a problem in quantum chemistry as many strongly-correlated systems are bond-breaking processes. Neuscamman has shown that size-consistency can be restored by introducing Jastrow factors on top of AGP.\cite{neuscamman:2012,neuscamman:2013,neuscamman:2016} Currently, Scuseria's group is developing wavefunctions with AGP as a mean-field.\cite{henderson:2019,khamoshi:2019,henderson:2020,khamoshi:2020,dutta:2020,khamoshi:2021,dutta:2021} This is trickier than for on-shell RG as there is no Hamiltonian whose eigenvectors are all AGP states. 

\subsection{Slater Determinant Expansion}
When the geminal coefficients do not depend on the geminal, i.e. $g^i_a \rightarrow g^i$, the Slater determinant expansion formula \eqref{eq:apig_slavnov_sd} reduces to
\begin{align}
\braket{h^M| g^M} = (M!)^2 \sum_{\{i\}} h^{\{i\}} g^{\{i\}} \label{eq:agp_esp_scalar}
\end{align}
since all of the permanents in the expansion simplify as each of their rows are identical. The summation is understood (as for the APIG case) as for collections of indices such that $i_1 < i_2 < \dots < i_M$, and $h^{\{i\}}g^{\{i\}}$ is a shorthand for $h^{i_1}g^{i_1}\dots h^{i_M}g^{i_M}$. The result is an elementary symmetric polynomial (ESP) in the variables $h^ig^i$. Scuseria's group has developed an algorithm to evaluate AGP matrix elements in terms of ESP with polynomial scaling.\cite{khamoshi:2019} It is not obvious why the scaling should be polynomial, but when written as a Sklyanin sum it will be immediately clear. The expressions for density matrix elements are explicitly
\begin{align}
\frac{1}{2}\braket{h^M| \hat{n}_k |g^M} &= (M!)^2 \sum_{\{i\}} h^{\{i\}} g^{\{i\}} \delta (k \in \{i\}) \label{eq:agp_esp_1rdm} \\
\frac{1}{4}\braket{h^M| \hat{n}_k \hat{n}_l |g^M} &= (M!)^2 \sum_{\{i\}} h^{\{i\}} g^{\{i\}} \delta (k,l \in \{i\}) \label{eq:agp_esp_Dij} \\
\braket{h^M|S^+_k S^-_l | g^M} &= (M!)^2  \frac{g^l}{g^k} \sum_{\{i\}} h^{\{i\}} g^{\{i\}} \delta (k \in \{i\}) \delta(l \notin \{i\}) \\
&= (M!)^2  h^k g^l \sum_{\{i\}'} h^{\{i\}'} g^{\{i\}'} \delta (k,l \notin \{i\}) \label{eq:agp_esp_Pij}
\end{align}
where the summation in the final line is over sets of $M-1$ elements without $k$ and $l$. This last formula is easier to manage.

Khamoshi et al.\cite{khamoshi:2019} have shown with an explicit recursion that all $p$-RDM elements are expressible in terms of the 1-RDM elements for any $p$. We will not reproduce their argument, but mention in the present case for the 2-RDM this yields explicitly
\begin{align}
D_{kl} &= \frac{g^l g^l}{g^l g^l - g^k g^k}\gamma_k - \frac{g^k g^k}{g^k g^k - g^l g^l} \gamma_l \\
P_{kl} &= \frac{g^k g^l}{g^k g^k - g^l g^l} \left(\gamma_k - \gamma_l \right).
\end{align}
These formulas may be seen by simple inclusion/exclusion arguments with the ESP defining $\gamma_k$: for $D_{kl}$, the terms containing both $k$ and $l$ survive, while for $P_{kl}$ the terms containing both $k$ and $l$ cancel. Similar expressions do not appear to hold for APIG, though even if they did it would not be so important. For AGP and APSG the expressions for the 2-RDM in terms of the 1-RDM are known. For RG states, it is not known if the 2-RDM is an explicit function of the 1-RDM, but both already require computing the same primitive elements (the partial derivatives $\frac{\partial \tilde{v}_a}{\partial \varepsilon_k}$) so there would no benefit. Of course for APIG itself the 1-RDM is already intractable to compute numerically.

\subsection{Sklyanin Sum}

As the geminals are identical, the contractions for AGP depend only on their rank. Specifically, we define the rank-$p$ contractions as
\begin{align}
\lambda (p) := \sum_i (h^i g^i)^p,
\end{align}
the \emph{power-sum symmetric functions} in the variables $h^i g^i$. The Sklyanin sum can then be evaluated in closed-form, with result
\begin{align} \label{eq:schur_det}
\braket{ h^M | g^M} &= M!
\begin{vmatrix}
\lambda (1) & 1 & 0 & 0 & \dots & 0 \\
\lambda (2) & \lambda (1) & 2 & 0 & \dots & 0 \\
\lambda (3) & \lambda (2) & \lambda (1) & 3 & \dots & 0  \\
\vdots \\
\lambda (M-1) & \lambda (M-2) & \lambda (M-3) & \lambda (M-4) & \dots & M-1 \\
\lambda (M) & \lambda (M-1) & \lambda (M-2) & \lambda (M-3) & \dots & \lambda (1)
\end{vmatrix}.
\end{align}
This result does not seem to be known in the AGP literature, though eventually we found it in the nuclear structure literature as number-projected coherent states.\cite{rowe:1991,chen:1995,otsuka:2001,rowe:2001,rowe_book,lu:2021} The determinant \eqref{eq:schur_det} is a representation of the Schur function of weight ($1^M$) over the $N$ variables $h^ig^i$. There are many other possible equivalent expressions in terms of other symmetric polynomials. The equivalence of the elementary symmetric polynomial \eqref{eq:agp_esp_scalar} and the above determinant of power-sum symmetric polynomials \eqref{eq:schur_det} is an identity known to Girard and Newton from the 17th century.\cite{macdonald_book}

The contractions $\lambda (p)$ may be pre-computed and the determinant \eqref{eq:schur_det} costs $\mathcal{O}(M^3)$. This approach is sub-optimal as we will see directly. Like the case for on-shell RG, the \emph{normalized} matrix elements could be computed from solutions of linear equations with Cramer's rule. Again this would be sub-optimal so we will not discuss it further.

The key property of the AGP scalar products is that they are recursive.\cite{rowe_book} Specifically, with the vacuum overlap $\braket{h^0|g^0}:= \braket{\theta|\theta} = 1$, and the Pochhammer symbols $\left( M \right)_p = \frac{M!}{(M-p)!}$
\begin{align} \label{eq:agp_sklyanin_scalar}
\braket{h^M | g^M} &= \sum_{p=1}^M (-1)^{p-1} (M)_p (M-1)_{p-1} \lambda (p) \braket{h^{M-p} | g^{M-p}}.
\end{align}
This recursive expression is exceptionally clean and straightforward to compute. The primitives $\lambda(p)$ can be precomputed with $\mathcal{O}(MN)$ cost, then the recursive expression can build the AGP overlap with a cost of $\mathcal{O}(M^2)$. Equations \eqref{eq:schur_det} and \eqref{eq:agp_sklyanin_scalar} have both been checked with the elementary symmetric polynomial result \eqref{eq:agp_esp_scalar}.

The density matrix elements are obtained from first 
\begin{align}
\frac{1}{2} \braket{h^M| \hat{n}_k |g^{M}} = g^k \frac{\partial}{\partial g^k} \braket{h^M|g^M}
\end{align}
and second derivatives (for $k \neq l$),
\begin{align}
\frac{1}{4} \braket{h^M | \hat{n}_k \hat{n}_l | g^M} &= g^k g^l \frac{\partial^2}{\partial g^k \partial g^l} \braket{h^M|g^M} \\
\braket{h^M|S^+_k S^-_l|g^M} &= g^l \frac{\partial}{\partial g^k} \braket{h^M|g^M} - g^l g^l \frac{\partial^2}{\partial g^k \partial g^l} \braket{h^M|g^M}
\end{align}
of the scalar product $\braket{h^M|g^M}$. The first derivatives can be constructed as
\begin{align}
\frac{\partial}{\partial g^k} \braket{h^M|g^M} &= \sum_{p=1}^M (-1)^{p-1}  (M)^2_p \frac{1}{p} \frac{\partial \lambda (p)}{\partial g^k} \braket{h^{M-p}|g^{M-p}}
\end{align}
with the derivatives of the $\lambda$ contractions
\begin{align}
\frac{\partial \lambda (p)}{\partial g^k} = p (h^k g^k)^{p-1} (h^k).
\end{align}
Second derivatives are obtained by iterating the first derivative expression
\begin{align}
\frac{\partial^2}{\partial g^k \partial g^l} \braket{h^M|g^M} 
&= \sum_{p=1}^M (-1)^{p-1} (M)^2_p \frac{1}{p} \frac{\partial \lambda (p)}{\partial g^l} \frac{\partial}{\partial g^k} \braket{h^{M-p}|g^{M-p}} \\
&= \sum_{p=1}^M \sum_{q=1}^{M-p} (-1)^{p+q} (M)^2_{p+q} \frac{1}{pq} \frac{\partial \lambda (p)}{\partial g^l} \frac{\partial \lambda (q)}{\partial g^k} \braket{h^{M-p-q}|g^{M-p-q}} 
\end{align}
in which we've used $(M)_p (M-p)_q = (M)_{p+q}$. Finally, the density matrix elements are computable with the same information as required for the scalar product
\begin{align} 
\frac{1}{2} \braket{h^M| \hat{n}_k |g^{M}} &= \sum^M_{p=1}  (-1)^{p-1} \left( M \right)^2_p \left( h^k g^k \right)^p \braket{h^{M-p}|g^{M-p}} \label{eq:agp_sklyanin_1rdm} \\
\frac{1}{4} \braket{h^M|\hat{n}_k \hat{n}_l|g^M} &= \sum^M_{p=1}\sum^{M-p}_{q=1} (-1)^{p+q} (M)^2_{p+q} \left(h^l g^l \right)^p \left( h^k g^k\right)^q \braket{h^{M-p-q}|g^{M-p-q}} \label{eq:agp_sklyanin_Dij} \\
\braket{h^M | S^+_k S^-_l | g^M} &= \frac{g^l}{g^k} \sum^M_{p=1} (-1)^{p-1} (M)^2_p \left(h^k g^k\right)^p \braket{h^{M-p}|g^{M-p}} \nonumber \\
&- \frac{g^l}{g^k} \sum^{M}_{p=1} \sum^{M-p}_{q=1} (-1)^{p+q} (M)^2_{p+q} \left(h^l g^l\right)^p \left(h^k g^k\right)^q \braket{h^{M-p-q}|g^{M-p-q}}. \label{eq:agp_sklyanin_Pij}
\end{align}
Equations \eqref{eq:agp_sklyanin_1rdm}, \eqref{eq:agp_sklyanin_Dij} and \eqref{eq:agp_sklyanin_Pij} have been checked numerically with equations \eqref{eq:agp_esp_1rdm}, \eqref{eq:agp_esp_Dij} and \eqref{eq:agp_esp_Pij}.

\section{APSG: Sparsity} \label{sec:apsg}
There are two other specific degenerate cases that we should mention. Both can be systematically corrected to be APIG in different circumstances. 
\subsection{Strongly-Orthogonal Geminals} \label{sec:APSG}
APSG has long been known to describe bond-breaking processes correctly.\cite{hurley:1953,kutzelnigg:1964,kutzelnigg:2010,kobayashi:2010,kutzelnigg:2012,surjan:2012,zoboki:2013,pernal:2014,jeszenszki:2014,pastorczak:2015,margocsy:2018,pernal:2018,pastorczak:2018,pastorczak:2019} Variational APSG is equivalent to the Piris natural orbital functional PNOF5,\cite{piris:2011,pernal:2013} meaning that the APSG 2-RDM is expressible directly in terms of its 1-RDM elements. Recently, the PNOF7 functional\cite{piris:2017,piris:2019,mitxelena:2020a,mitxelena:2020b,rodriguez:2021,piris:2021} has shown convincing numerical results. In its simplest form PNOF7 is based on closed-shell pairs (generalizations are possible and pointed out in ref \cite{piris:2017}), but the intra-pair and inter-pair interactions are treated separately. The intra-pair terms are essentially the same as APSG, while the inter-pair terms look very similar to AGP (see in particular, equation (4.49) on page 153 of ref \cite{coleman_book}). We tried but could not find a case of APIG which reduced to PNOF7, though we are not surprised as Piris' development of PNOF7 was based on ensemble (rather than pure state) 2-RDM N-representability arguments. Another promising natural orbital functional based on the entropy has recently been published,\cite{wang:2022} and likewise appears to be ensemble N-representable. While not strictly based on natural orbitals, other seniority-zero functionals have been developed.\cite{vu:2020}

APSG is a geminal product for which the coefficients belong to disjoint subspaces, i.e. each primitive $S^+_i$ only contributes to one geminal in the product:
\begin{align}
\ket{ \{g\}} = \prod_a \left( \sum_{i\in \Omega_a} g^i_a S^+_i \right) \ket{\theta}
\end{align}
where $\Omega_a$ denotes the set of spatial orbitals associated with the geminal $G^+(g_a)$. Strong orthogonality means that for $a \neq b$ the sets $\Omega_a$ and $\Omega_b$ are disjoint and so for each spatial orbital $i$, there is only one non-zero coefficient $g^i_a$. Lower indices of the geminal coefficients could be suppressed, but will be kept to be more coherent. Each spatial orbital occurs in a single $\Omega$. The $\Gamma$-contractions become particularly simple in this case. For two APSG states, $\ket{\{h\}}$ and $\ket{\{g\}}$, we will assume that the distribution of spatial orbitals is identical and that the geminals have been ordered in the same manner. The \emph{only} terms that do not vanish are the rank-1 diagonal contractions
\begin{align}
\Gamma (h_a, g_a) = \sum_{i \in \Omega_a} h^i_a g^i_a.
\end{align}
This further simplifies to sums over two elements for the Generalized Valence Bond/Perfect-Pairing (GVB/PP) wavefunction,\cite{goddard:1967,hay:1972,hunt:1972,goddard:1973,goddard:1978} which is an APSG in which each geminal consists of exactly two spatial orbitals. 

Off-diagonal rank-1 contractions $\Gamma (h_a,g_b)$ vanish since the non-zero coefficients $h^i_a$ are disjoint from the non-zero coefficients $g^i_b$. All higher rank $\Gamma$-contractions vanish identically. For two APSG states, the scalar product reduces to one summand:
\begin{align}
\braket{ \{h\} | \{g\}} = \prod_a \Gamma( h_a,g_a).
\end{align}
The 1-pair form factors are only non-zero if $k \in \Omega_a$, hence for each $k$ there is only one non-zero form factor
\begin{align}
\braket{\{h\} | S^+_k | \{g\}_a} = h^k_a \; \delta (k \in \Omega_a) \prod_{c\neq a} \Gamma(h_c,g_c).
\end{align}
The 2-pair form factors are likewise only non-zero if $k\in \Omega_a$ \emph{and} $l \in \Omega_b$, while it is also understood that $a\neq b$
\begin{align}
\braket{ \{h\} | S^+_k S^+_l | \{g\}_{a,b}} = h^k_a h^l_b \; \delta (k\in \Omega_a) \delta (l \in \Omega_b) \prod_{c\neq a,b} \Gamma(h_c,g_c).
\end{align}
The density matrix elements each have one non-zero term. The diagonal-correlation elements are
\begin{align}
\frac{1}{2} \braket{\{h\} | \hat{n}_k | \{g\}} &= h^k_a g^k_a \;\delta(k\in \Omega_a) \prod_{c \neq a} \Gamma (h_c,g_c) \\
\frac{1}{4} \braket{\{h\} | \hat{n}_k \hat{n}_l | \{g\}} &= h^k_a g^k_a h^l_b g^l_b \;\delta(k\in \Omega_a) \delta (l \in \Omega_b) \prod_{c \neq a,b} \Gamma (h_c,g_c)
\end{align}
while the pair-correlation elements are only non-zero if both $k$ and $l$ belong to the same set $\Omega_a$
\begin{align}
\braket{ \{h\} | S^+_k S^-_l | \{g\}} = h^k_a g^l_a  \; \delta (k,l \in \Omega_a) \prod_{c\neq a} \Gamma(h_c,g_c).
\end{align}
The double sum that contributes to the pair-correlation function here is identically zero. Proper normalization reduces the 2-RDM elements to the form usually quoted
\begin{align}
D_{kl} &= \gamma_k \gamma_l\; \delta (k \in \Omega_a) \delta (l \in \Omega_b) \\
P_{kl} &= \sqrt{\gamma_k \gamma_l} \; \delta (k,l \in \Omega_a).
\end{align}

APSG is variationally feasible as strong orthogonality ensures that higher-rank contractions vanish identically so that the Sklyanin sum has a small number of terms to evaluate. In particular, only the diagonal rank-1 terms survive. Weaker orthogonality criteria can be enforced to include more non-zero terms in the resulting Sklyanin sum. In particular, if any two sets $\Omega_a$ and $\Omega_b$ share one common element $m$, then the off-diagonal rank-1 $\Gamma(h_a,g_b) = h^m_a g^m_b$, the diagonal rank-2 $\Gamma(h_a,h_b,g_a,g_b)=h^m_a h^m_b g^m_a g^m_b$ but all off-diagonal rank-2 and higher-rank contractions remain zero. Allowing more elements to be shared between different geminals systematically includes more terms in the Sklyanin sum, and thus APSG can be systematically corrected to APIG. While not quite synonymous with $p$-orthogonality \cite{wilson:1976,cassam:2006,cassam:2010,cassam:2012}, one could say that two APSG type geminals are $p$-orthogonal if they share at most $p-1$ elements. Limacher\cite{limacher:2016} has considered a similar approach by projecting APIG against APSG vectors in a coupled-cluster-like manner.

\subsection{AP1roG/pCCD}
AP1roG\cite{limacher:2013} is a geminal wavefunction in which the orbitals have been separated into occupieds and virtuals. Loosely speaking, the occupieds are strongly-orthogonal while the virtuals are weakly-orthogonal. As such, it presupposes that one Slater determinant is a reasonable first approximation. Each geminal has a contribution from one occupied and each virtual
\begin{align}
\ket{\{g\}} = \prod_a \left( S^+_a + \sum_{i\in virt} g^i_a S^+_i \right) \ket{\theta},
\end{align}
and since each $S^+$ can only occur once, this is equivalent to pCCD:\cite{stein:2014}
\begin{align}
\ket{\text{pCCD}} = \exp \left( \sum_{\substack{i \in occ \\ v \in virt }} t^v_i S^+_v S^-_i \right) \ket{\text{HF}}.
\end{align}
The rank-1 contractions are
\begin{align}
\Gamma(h_a,g_b) = \delta_{ab} + \sum_{i \in virt}h^i_a g^i_b
\end{align}
while all higher-rank contractions are strictly the sums over products of virtual coefficients. As a result, AP1roG / pCCD is \emph{not} feasible variationally. It is instead solved by projection on one particular Slater determinant along with its corresponding pair double-excitations. It is feasible, cheap even, as the few permanents that are present are limited in size to $2\times 2$. If, in a Slater determinant basis, APIG has one dominant contribution, then AP1roG / pCCD is the best first approximation with successive approximations described by ratios of determinants of cluster amplitudes.\cite{zhao:2016,fecteau:2021} Others have considered similar wavefunction forms based on ratios of determinants as well.\cite{kim:2021} In terms of the ground-state energy, AP1roG/pCCD describes many strongly-correlated systems quite well.\cite{limacher:2013,limacher:2014a,limacher:2014b,henderson:2014a,henderson:2014b,boguslawski:2014a,boguslawski:2014b,boguslawski:2014c,tecmer:2014} In weakly-correlated regimes, the physical wavefunction is adequately described as a HF mean-field plus pair doubles. In bond-breaking regimes, the physical wavefunction will tend towards being APSG. Both limits are in the scope of AP1roG/pCCD. There are however drawbacks. Solving by projection means that properties other than the energy are suspect. Systematic corrections are difficult to define as the Hilbert space isn't easily described by a set of orthogonal AP1roG/pCCD vectors. However, AP1roG/pCCD remains the method to beat. Results building upon AP1roG/pCCD are quite promising.\cite{boguslawski:2015,degroote:2016,boguslawski:2016a,boguslawski:2016b,boguslawski:2017,boguslawski:2019,nowak:2019,nowak:2021,boguslawski:2021,marie:2021,baran:2021} 

\section{Numerical Results} \label{sec:numbers}
The goal is to judge whether the incorrect behaviour shown by the RG mean-field in ref\cite{johnson:2020} is fixable by off-shell RG states or by APIG. Variational calculations for both off-shell RG and APIG were performed for H$_4$, H$_6$ and H$_8$, in the basis of OO-DOCI orbitals (STO-6G) computed with GAMESS (US)\cite{barca:2020} for ref.\cite{johnson:2020} As off-shell RG and APIG both scale intractably, it is not important for the implementation or the numerical optimization to be efficient. Geminal coefficients were pre-conditioned with the covariance matrix adaptation evolution strategy (CMA-ES)\cite{hansen:2001} before being optimized with the Nelder-Mead simplex algorithm.\cite{nelder:1965} Full configuration interaction (FCI) results were computed with psi4\cite{sherill:1999,parrish:2017} also for ref.\cite{johnson:2020}

\begin{figure}[ht!]
	\begin{subfigure}{\textwidth}
		\includegraphics[width=0.475\textwidth]{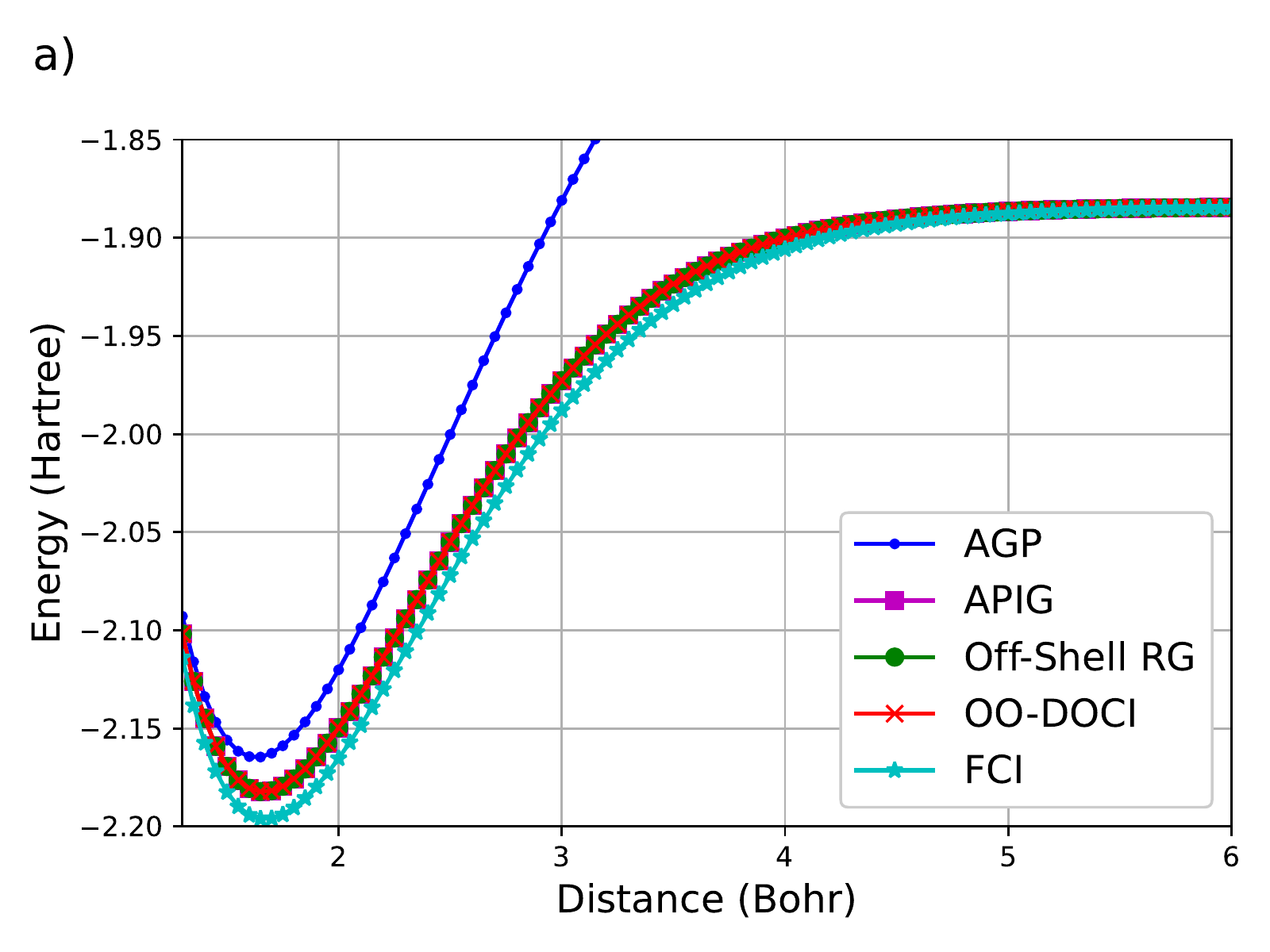} \hfill
		\includegraphics[width=0.475\textwidth]{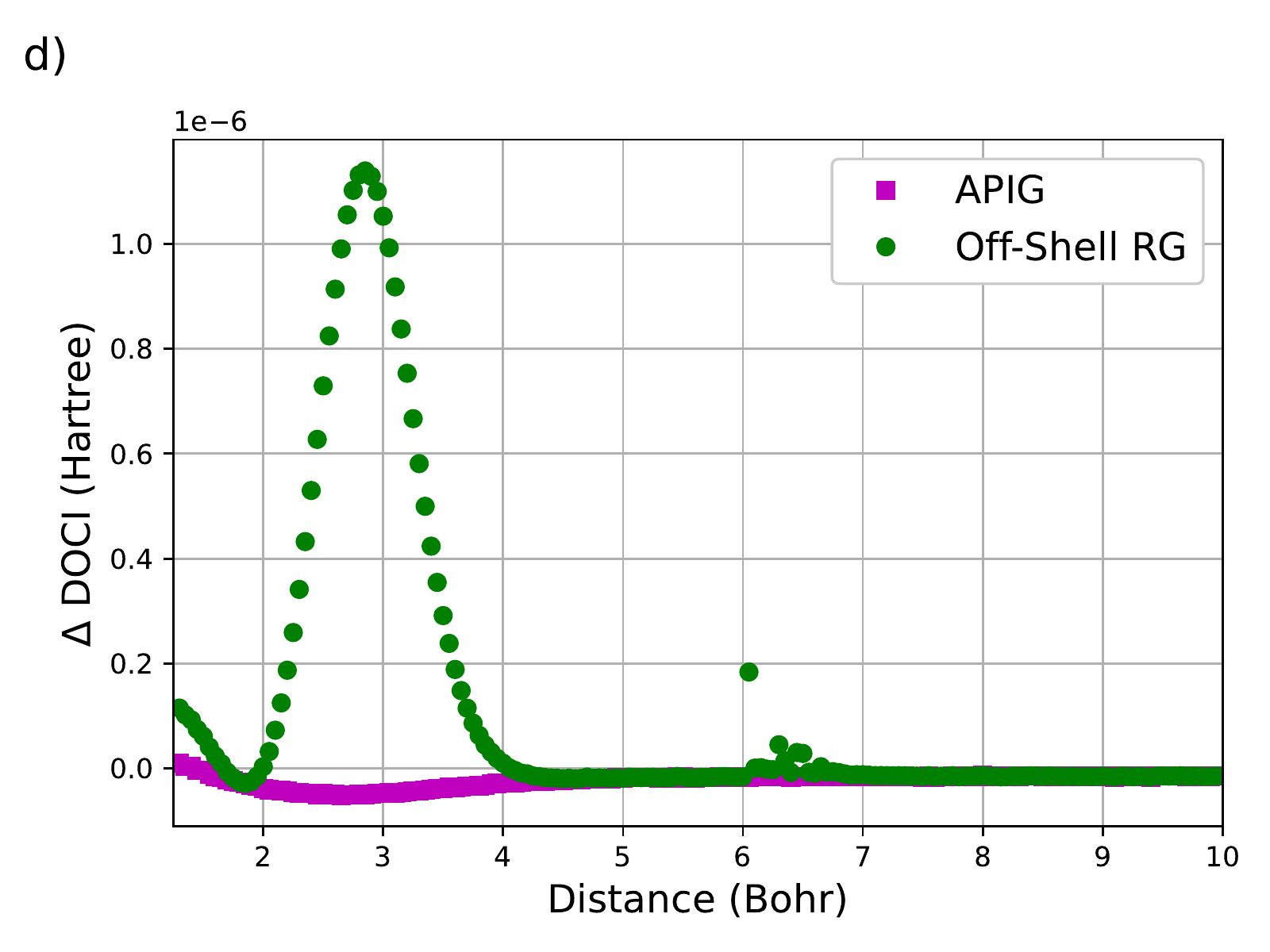}
	\end{subfigure}
	\begin{subfigure}{\textwidth}
		\includegraphics[width=0.475\textwidth]{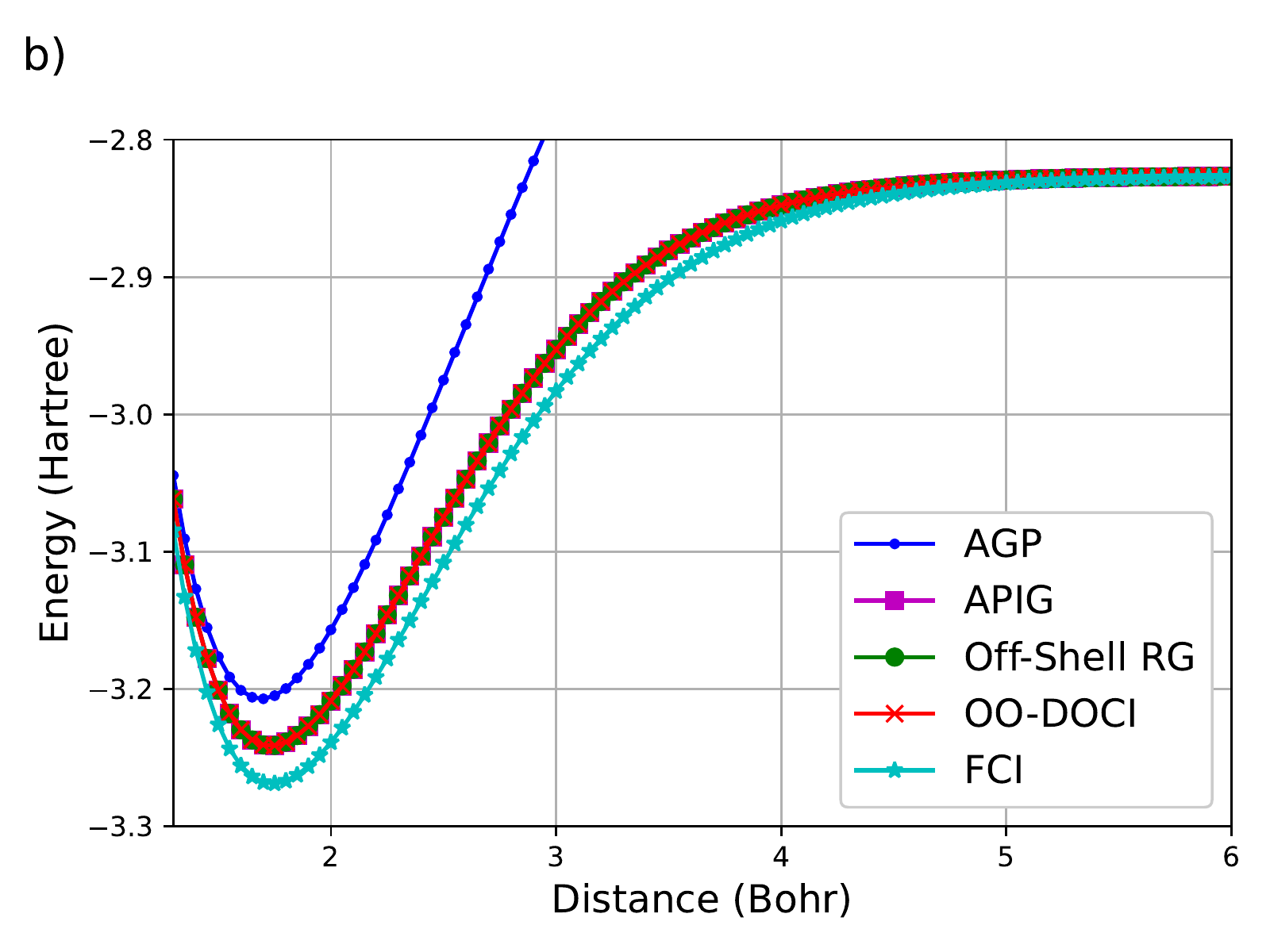} \hfill
		\includegraphics[width=0.475\textwidth]{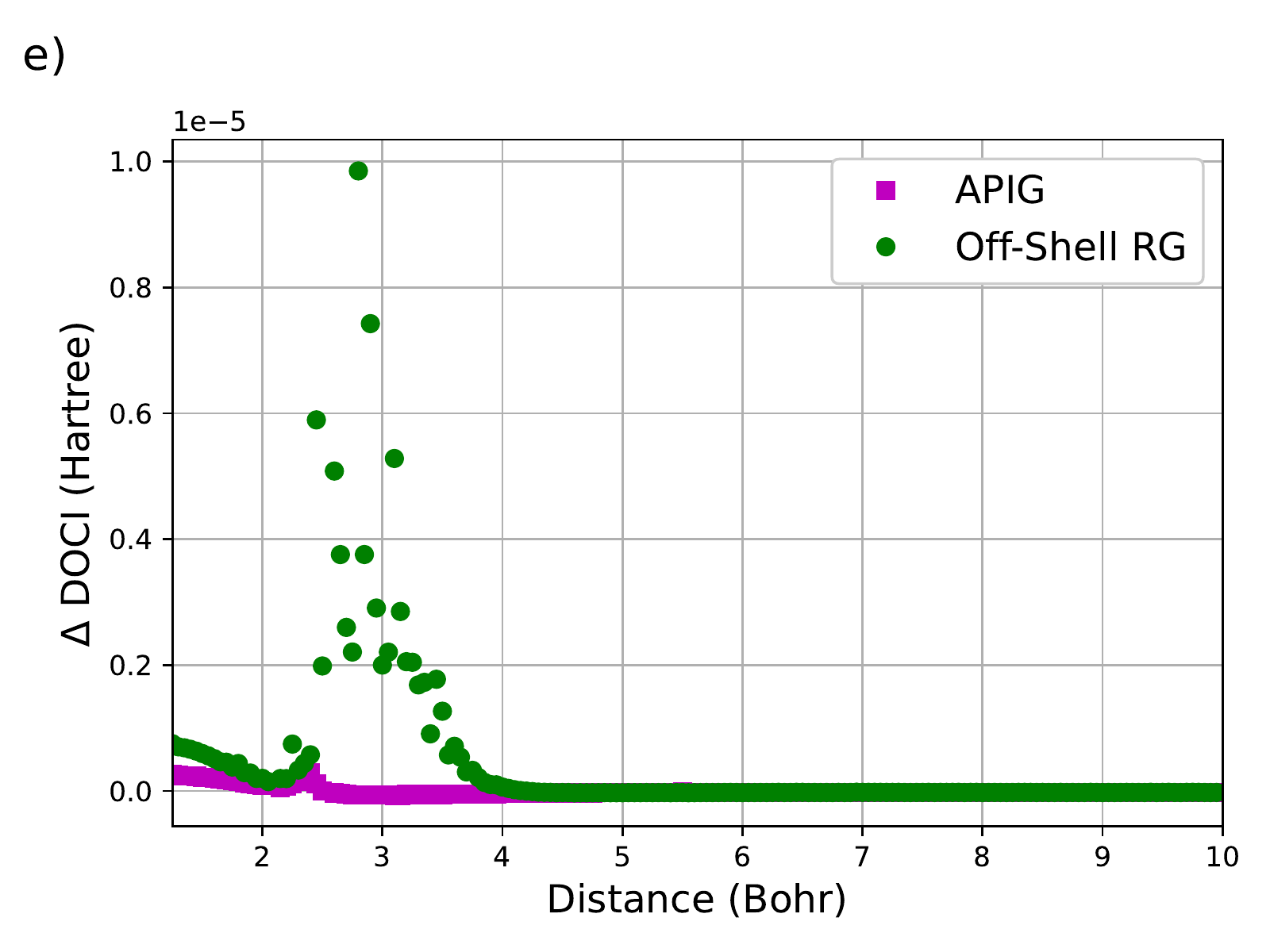}
	\end{subfigure}
	\begin{subfigure}{\textwidth}
		\includegraphics[width=0.475\textwidth]{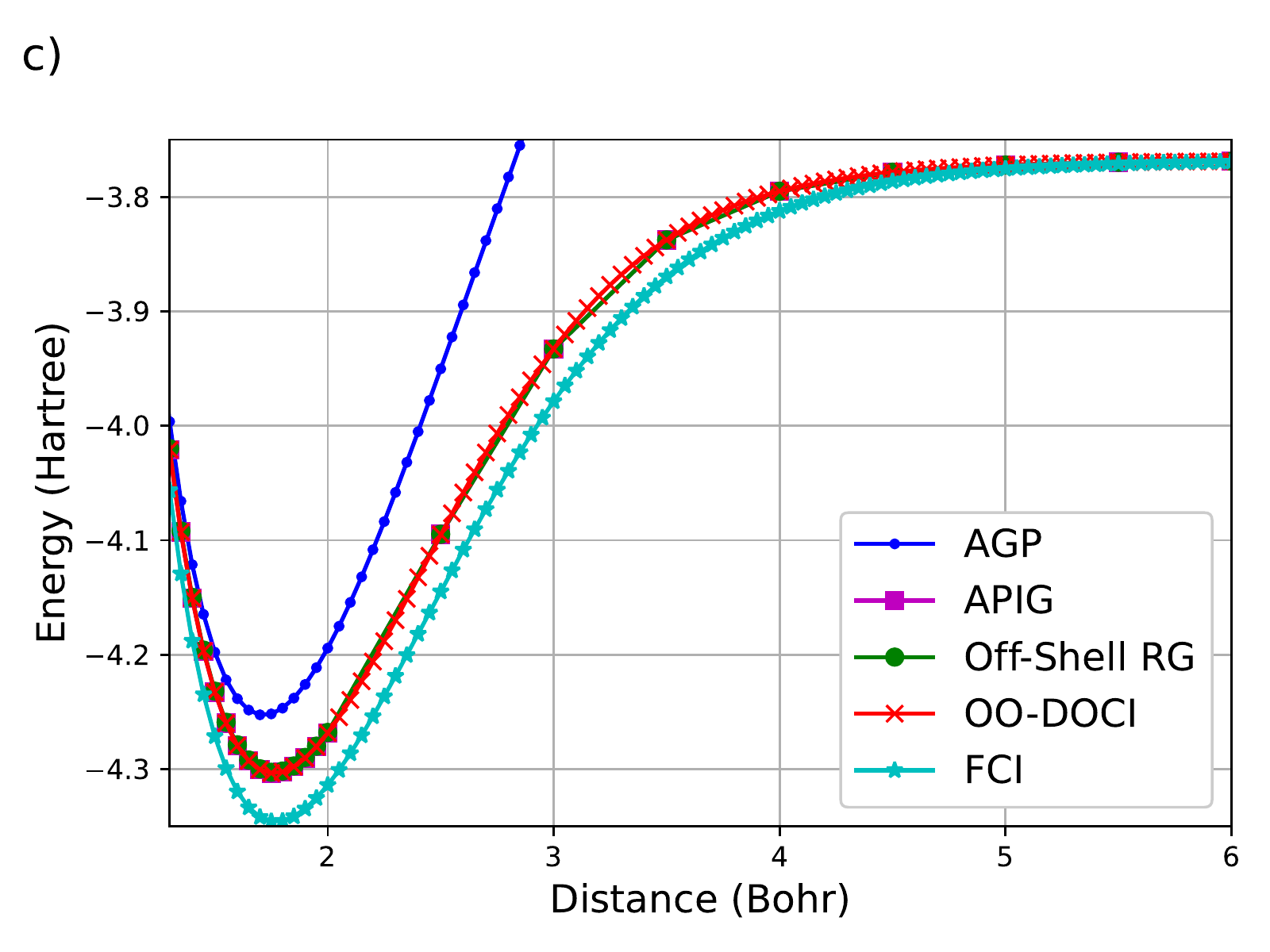} \hfill
		\includegraphics[width=0.475\textwidth]{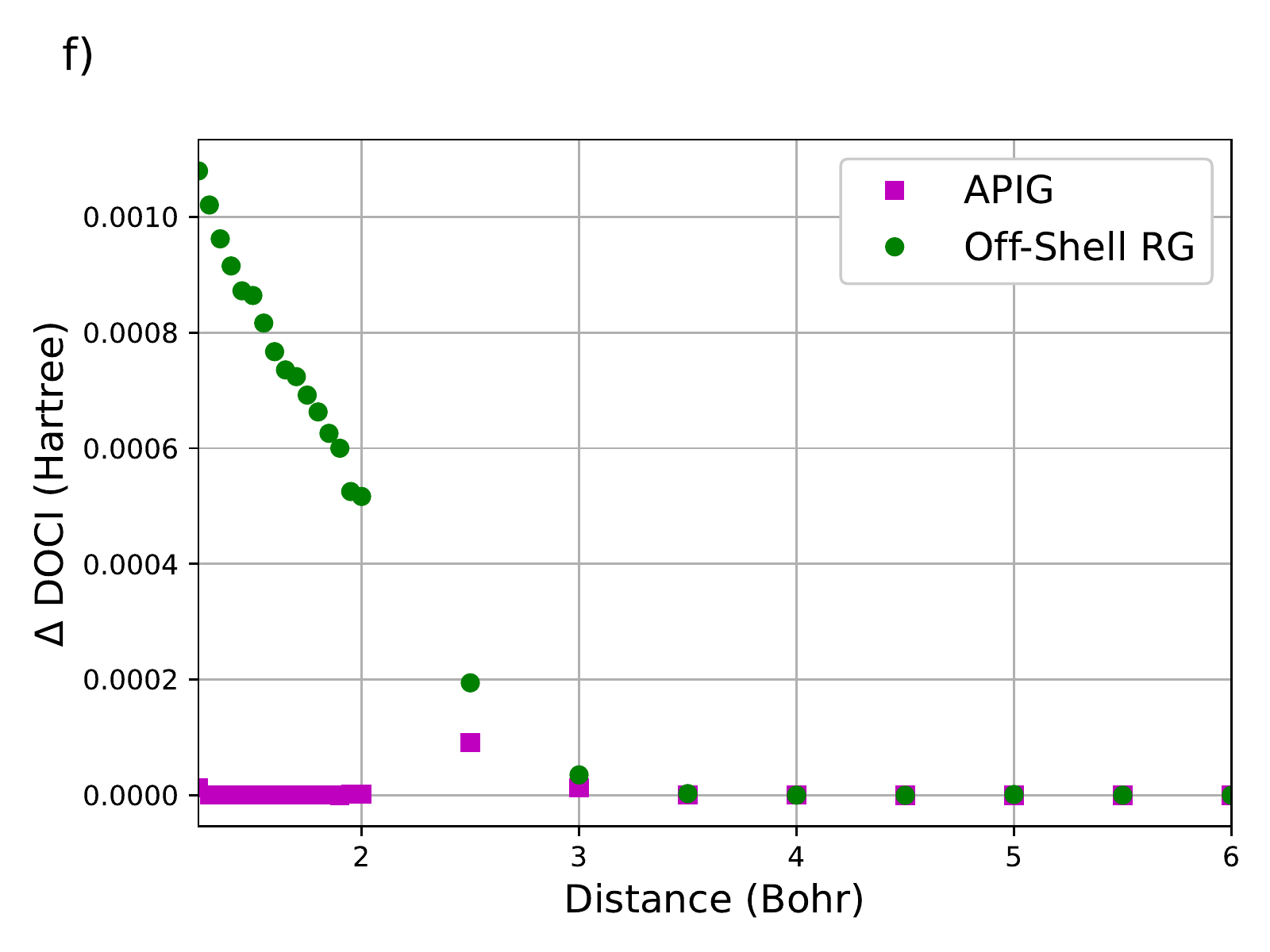}
	\end{subfigure}
		\caption{(a)-(c) Bond dissociation curves for H$_4$, H$_6$, and H$_8$ computed with AGP, off-shell RG, APIG, OO-DOCI, and FCI. (d)-(f) Errors for off-shell RG and APIG with respect to OO-DOCI. Results were all computed with the STO-6G basis in the OO-DOCI optimized orbitals. OO-DOCI and FCI results are from ref.\cite{johnson:2020}}
		\label{fig:H_curves}
\end{figure}
Variational curves obtained for the dissociation of symmetric hydrogen chains are presented in figure \ref{fig:H_curves}. AGP results are presented mainly for completeness. AGP is not size-consistent and does not treat bond dissociation processes well. Both off-shell RG and APIG are very close to the OO-DOCI results. For H$_4$ and H$_6$ there is structure in the deviation of off-shell RG from OO-DOCI, though in both cases the errors are very small. For H$_8$, the deviation of off-shell from OO-DOCI is larger, though this is because the convergence criterion for our solver needed to be loosened. The calculations quickly become very heavy and therefore the H$_8$ curve has fewer points. In all cases, the deviation of off-shell RG from OO-DOCI is maximal near the minimum, where the system is weakly-correlated. It is clear that the non-physical avoided crossing observed for the on-shell RG ground state is not replicated with off-shell RG. Thus, the failure in ref.\cite{johnson:2020} is \emph{not} a feature of the RG geminal form, but of the RG state used. It is reasonable to presume that another RG state could do better, and we will show this definitively in an upcoming contribution. The APIG results are numerically identical to OO-DOCI. APIG appears to go below OO-DOCI which should be physically impossible. To the precision that we can trust both results, APIG and OO-DOCI are identical.

\section{Conclusion}
Scalar products and density matrix elements have been computed for APIG directly in the basis of the pairs. The result, the Sklyanin sum, is a sum over all possible ways of contracting the geminal coefficients, which is analogous to Wick's theorem for fermions or bosons. For APIG the Sklyanin sum is intractable though degenerate cases show how it may be made feasible in three distinct ways. For RG states, the rational structure of the geminal coefficients reduces the rank of each contraction to one, and Richardson's equations lead to all remaining terms, except one, cancelling out. For AGP, the geminals are identical so the contractions depend only on their rank. This leads to a simple clean recursion. For APSG only a small number of contractions are non-zero since the geminals are strongly-orthogonal. This may be relaxed systematically to eventually become APIG.

Variational bond dissociation curves for symmetric hydrogen-chain dissociations were calculated with both off-shell RG and APIG. This was to establish whether the un-physical results of ref.\cite{johnson:2020} were due to a problem with the RG ground state, the RG geminal form, or of the APIG approximation to DOCI. The present results show that both off-shell RG and APIG give virtually the same energy as DOCI, so the problem must lie with the RG ground state. This is fixable with other RG states, which we will demonstrate in an upcoming contribution. 

\section{Acknowledgements}
P.A.J. was supported by NSERC and Compute Canada. C.-\'{E}.F. is grateful for funding from the Vanier Canada Graduate Scholarships.

\bibliography{APIG_arXiv}

\bibliographystyle{unsrt}

\end{document}